\documentclass[pre,twocolumn,english,superscriptaddress,floatfix]{revtex4}
\usepackage{graphicx}
\usepackage{amsmath}
\usepackage{amssymb}
\usepackage{color}
\makeatletter

\usepackage{babel}
\usepackage{algorithm}
\usepackage{algorithmic}
\newfloat{mybox}{thp}{lp}
\floatname{mybox}{Algorithm}

\newcommand\Pf{\operatorname{Pf}}
\newcommand\figref[1]{Fig.~\ref{#1}}
\newcommand\myeqref[1]{Eq.~(\ref{#1})}

\makeatother
\begin{document}

\title{Numerically exact correlations and sampling in the two-dimensional Ising spin glass}

\author{Creighton K. Thomas}
\affiliation{Dept. of Materials Science and Engineering, Northwestern University, Evanston, Illinois 60208-3108, USA}
\author{A. Alan Middleton}
\affiliation{Department of Physics, Syracuse University, Syracuse, New York 13244, USA}

\begin{abstract}
A powerful existing technique for evaluating statistical
mechanical quantities in two-dimensional Ising models is based on constructing a matrix 
representing the nearest neighbor spin couplings and then evaluating the Pfaffian of the matrix.
Utilizing this technique and other more recent
developments in evaluating elements of inverse matrices and exact sampling, a method and computer code
for studying two-dimensional Ising models is developed. The formulation of this method is convenient and
fast for computing the partition function and spin correlations. It is also useful for exact sampling,
where configurations are directly generated with probability given by the
Boltzmann distribution. These methods apply to Ising model samples with arbitrary nearest-neighbor couplings and
can also be applied to general dimer models.
Example results of computations are described, including comparisons with analytic results for the ferromagnetic
Ising model, and timing information is provided.
\end{abstract}
\maketitle

\section{Introduction}

Just over 50 years ago, Kasteleyn \cite{KasteleynDimers} and Fisher and Temperley \cite{FisherDimers}
presented analytic combinatorial methods for counting dimer packings on a lattice;
these techniques were soon applied \cite{KasteleynIsing,FisherIsing} to computing the partition function
for the pure ferromagnetic two-dimensional Ising model \cite{Robertson,KrauthBook}. These methods continue to
be extended and improved to study two-dimensional models in statistical mechanics. Such methods
were then extended \cite{SaulKardar} to numerically compute the thermodynamic properties of disordered magnets.
They have allowed for a precise and extensive study of the statistical mechanics of disordered models
\cite{GLV,JorgEtAl,ThomasHuseMiddleton}. This paper presents a detailed description of 
a numerical approach to implement these combinatorial techniques.

To review the power of these techniques in more detail, consider an Ising model where spins on a planar lattice can
take on one of two values and the energy is given by the sum over possible ferromagnetic or antiferromagnetic
interactions between pairs of neighboring spins.
Directly evaluating the partition function of this model with $N$ spins involves a sum of Boltzmann
factors over the $2^N$ spin configurations.
But combinatorial techniques allow for a much more compact evaluation. The Ising configurations can be put into correspondence with
dimer coverings on a related lattice, where dimer coverings are choices of edges so that each node of the lattice is in 
exactly one chosen edge. Weighted sums $Z_d$ over all dimer coverings give the partition function $Z$ for the Ising problem, with $Z=Z_d$.
The sum over all dimer coverings can in turn be expressed as the Pfaffian \cite{Robertson,KrauthBook} of a weighted and signed
adjacency-like matrix, the Kasteleyn matrix \cite{KasteleynDimers}; for a skew-symmetric matrix, the square of its Pfaffian 
is equal to its determinant.
For the specific case of regular lattices and interactions, the Pfaffians can even be evaluated analytically by
direct diagonalization of the Kasteleyn matrix \cite{KasteleynIsing,FisherIsing},
allowing for exact evaluation of thermodynamic quantities and studies of phase transitions.
Pfaffians (or also determinants) of $m\times m$ matrices can be defined directly as the
sum over permutations whose number grows exponentially with $m$,
but the matrix can be be simplified by column and row eliminations, so that
the Pfaffian (or determinant) can be evaluated in time polynomial in $m$. As the Kasteleyn matrix used
here is of size $4N\times 4N$, $Z$ can be found for general planar
Ising models in time polynomial in $N$. (Note that as the arithmetic precision needed for
a stable calculation of $Z$ depends on $\beta$, there is a $\beta$ dependent prefactor for the running
time which scales roughly as $\beta$ \cite{ThomasMiddletonSampling}.)

This technique was subsequently generalized
to include inhomogeneous couplings between nearest neighboring Ising spins by Saul
and Kardar \cite{SaulKardar} for numerical work. In the Ising spin glass, the nearest neighbor
interactions can be either ferromagnetic or antiferromagnetic. For a given random choice of couplings of
arbitrary sign, the Pfaffian of the Kasteleyn matrix can be computed and used to derive
thermodynamic potentials and susceptibilities. If the couplings are of fixed magnitude $J$ but random sign (the bimodal
distribution), the exact dependence of $Z$ on inverse temperature $\beta$ can be written as a polynomial in $e^{-2\beta J}$ \cite{SaulKardar}.
More generally, this numerical approach has allowed for a detailed study of the thermodynamics of the 2D Ising spin glass, even
with continuous disorder distributions.
One example of the more important recent developments in these algorithms has been the application
of nested dissection and integer arithmetic \cite{GLV} for computing $Z(T)$ for the bimodal distribution in larger systems.
Applying Wilson's dimer sampling technique \cite{Wilson}, this numerically exact approach has also been
used to generate random samples of configurations of random Ising models \cite{ThomasMiddletonSampling}.
This sampling method bypasses the long equilibration times that arise in Markov Chain Monte Carlo methods.

In this paper, we describe a version of nested dissection as applied to the
Pfaffian techniques, with the goal of simplifying and extending the calculation of correlation functions and
sample configurations.
For a two-dimensional Ising sample with arbitrary nearest-neighbor couplings,
we describe how this technique can be used to compute the partition function,
to calculate correlation functions, and to randomly choose sample spin configurations.
We find that, due to near cancellations of intermediate sums,
multiple precision floating point arithmetic is needed to find accurate results at temperatures of interest
for system sizes of size about $20^2$ or larger (though the sets of integer
fields used in Ref.\ \onlinecite{GLV} could also be used for the partition function in the bimodal case).
We note that correlation functions were computed at $T=0$
by Blackman and Poulter for the bimodal case \cite{PoulterBlackman} using a different approach.
We simplify the sampling technique used in
our previous work \cite{ThomasMiddletonSampling} by simplifying the matrices and
by using a different approach to maintain computed correlation functions as spins are sampled. Many of these
improvements are based on the FIND (fast inverse using nested dissection) algorithm \cite{FIND} which computes desired elements
of a matrix inverse quickly and was developed to compute nonequilibrium Green's function
applications in nanodevices.
This particular flavor of hierarchical decomposition 
is very well suited to the geometry of the mapping between two-dimensional Ising models and dimer coverings.
While much of this algorithm is implicit in previous work, we assemble these methods into a form adapted to studying
the statistical mechanics of the Ising model, with novel applications to computing correlation functions,
and emphasize the nature of the algorithms as a renormalization procedure and clarify the sampling procedure.
This formulation is also significantly faster in practice.
We present
comparisons with analytic results, sample results for the spin glass case, and empirical results for the timings.
A version of the computer code for computing partition functions, written in C++, is available in the supplemental materials for this paper at [publisher URL]
or by download \cite{code}.
Extensions of this version of the code have been checked against analytic predictions for correlation functions in the ferromagnet and
against other exact codes for small spin glass samples. This code can be used to study pure, random bond, and spin glass models.

\section{Ising model, dimers, \& Pfaffian}

In this section, we state the standard Ising spin glass Hamiltonian and recall the mapping between Ising spin configurations
and dimer coverings \cite{KasteleynIsing,Barahona,Robertson,KrauthBook}.
We also review the definition of the Pfaffian of the Kasteleyn matrix and its relation to the partition function
of the Ising model.

A state $S$ of the Ising model in two dimensions on a rectangular sample composed of
$L_x \times L_y$ spin variables $s_i$ is given by a choice for each $s_i$, where each $s_i$
is restricted to $s_i=\pm 1$. 
The $n=L_x \times L_y$ sites $i$ lie on a square grid. There are $2^n$ possible spin
configurations in the state space $\mathcal{S}$.
The statistical mechanics of this model is governed by the standard Hamiltonian
\begin{eqnarray}
\mathcal{H}(S) &=& -\sum_{\langle ij \rangle}J_{ij} s_i s_j,
\end{eqnarray}
where the sample-dependent bond strengths $J_{ij}$ are quenched, i.e., fixed
in time, and connect nearest neighbor spin pairs $\langle ij \rangle$.
The spins lie on the nodes of a graph $G$ whose edges $\langle i j \rangle$
connect these nearest neighbor pairs.
For open or free boundary conditions or for fixed spins on the boundaries, these pairs form the edges of a planar graph.
For periodic boundary conditions, nearest neighbor pairs $\langle i j \rangle$ include bonds that wrap the sample around
a torus by connecting the top row
of the array to the bottom row and the right column to the left column.
In equilibrium at temperature $T=\beta^{-1}$, the probability
$P(S)$ of a spin state $S$ is $P(S)=\exp^{-\beta\mathcal{H}(S)}/Z$, where the partition function
is $Z=\sum_{S\in\mathcal{S}}\exp^{-\beta\mathcal{H}(S)}$. Numerical
derivatives of $Z(T)$
with respect to $T$ allow for the computation of energy $E(T)$, the entropy $S(T)$, and the
heat capacity $C(T)$. Exact sampling will be
taken to mean that configurations $S$ are generated with the correct probability $P(S)$, within numerical accuracy.
The correlation functions that will be computed by the algorithm are spin-spin correlation functions,
$\langle s_i s_j \rangle$, where the average is taken over all configurations weighted by their
equilibrium probability, i.e., $\langle s_i s_j\rangle = \sum_S P(S) s_i s_j$. Computing these correlations allows for a direct measure of the correlation length
and the density of relative domain walls.
Though we describe the techniques using square lattice samples with open boundaries or
with periodic boundaries, the techniques presented
generically apply to arbitrary graphs on low-genus surfaces \cite{KasteleynDimers,GLV}.

A given spin configuration $S$ in the Ising model can be represented by a set of relative domain walls and by the value of 
a single spin. These domain walls can be defined relative to any reference spin configuration $S^r$;
one simple choice for $S^r$ is
the fixed direction configuration $S^+$, with all $s_i=+1$. This is the choice that we will use in this paper.
(Another example choice would
be  a ground state configuration $S^{gs}$ that minimizes $\mathcal{H}$.)
The domain walls divide the spins into connected
sets of spins that are either all aligned with or all opposite to the spins in $S^r$. 
These domain walls can be drawn as loops on the dual graph $G_D$. The graph $G_D$ has nodes at the center
of each (square) plaquette of $G$. The edges of $G_D$ are dual to the edges in $G$: they are in one-to-one correspondence,
with each edge in $G_D$ crossing one edge in $G$.
Given an arbitrary spin configuration
$S$ and a nearest neighbor pair of spins $\langle ij \rangle$,
the dual edge that crosses the bond connecting $i$ to $j$ is in a domain wall if $s_i s_j \ne s^r_i s^r_j$. For
the choice $S^r=S^+$, the domain walls separate up spins from down spins. As the domain walls
are closed loops, an even number of domain wall segments meet at each node in $G_D$.

\begin{figure}[h]
\centering
\includegraphics[width=3.0in]{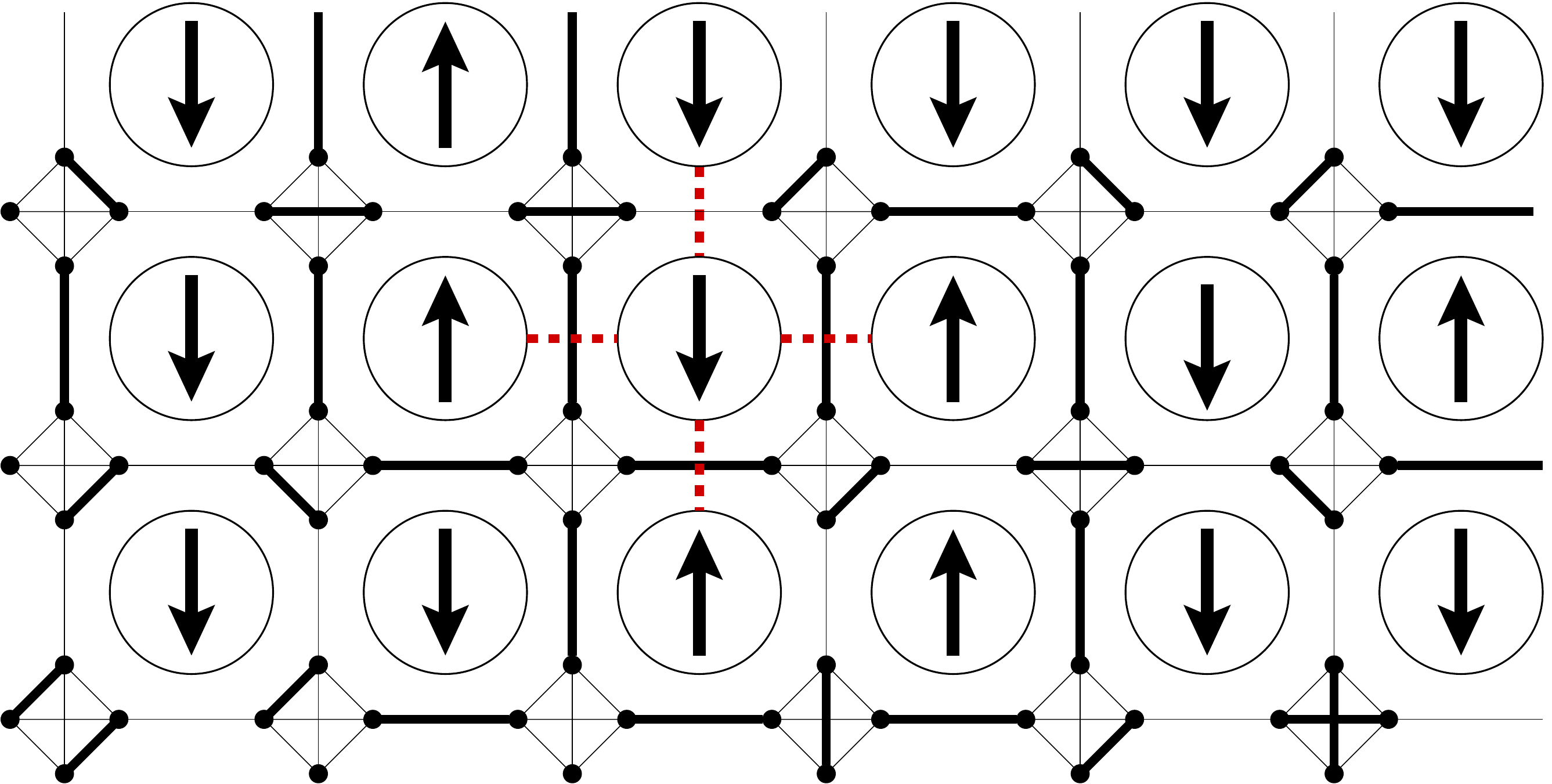}
\caption{(color online).
Correspondence between spin state configurations and complete dimer coverings on the decorated dual graph $G_D^*$
for a periodic spin lattice of size $L_x\times L_y = 6\times 3$. A sample configuration of Ising spins $s_i=\pm 1$ are
represented by the arrows inside the large circles. These spins are coupled by horizontal and vertical bonds 
of strength $J_{ij}$. The (red) dashed lines indicate the bonds for one example spin. The spins and bonds are the vertices and
edges, respectively, of the Ising model graph $G$.
The nodes of the decorated dual graph $G_D^*$ are drawn as small circles and the edges are indicated by the thin
and thick solid lines.
The edges of $G_D^*$ are either internal to a Kasteleyn city (the sets of 4 fully connected nodes) or connect neighboring
cities. Those that are internal to a city have a weight of $1$ while those connecting cities have a weight $w=\exp(-2\beta J_{ij})$,
where the $J_{ij}$ is the coupling strength of the bond crossing the dual edge.
An example of a complete dimer covering $M$
corresponding to the displayed spin configuration is indicated by the heavy lines: such a choice of edges
includes all nodes in $G_D^*$ exactly once.
The intercity edges belonging to the covering $M$ separate the up spins from the down spins and so
compose the relative domain walls (here we are assuming that the reference configuration $S^r$ is the configuration with
all spins up, $s_i = +1$). Note that for any Kasteleyn city surrounded by 4 spins of identical sign, there are 3
ways to arrange the dimers on that city. Two examples of these arrangements can be seen in the lower left and lower right cities.
In other cases, the choice of Kasteleyn city edges is uniquely determined by the domain walls.
}
\label{fig:dimers}
\end{figure}

The configurations in the Ising model may be put into correspondence with a complete dimer covering
problem on a decorated dual graph $G_D^*$. For the case we are considering, where $G$ is a square grid,
each node of $G_D$ can be replaced by a Kasteleyn city \cite{KasteleynIsing}, which is a subgraph composed of four
fully connected nodes.
(Note that a Kasteleyn city can be found from a Fisher city \cite{FisherIsing} by
Pfaffian elimination.) By replacing each lattice point in the dual $G_D$ with a Kasteleyn city,
one arrives at the decorated dual graph $G_D^*$ shown in \figref{fig:dimers}.
This larger graph allows for a correspondence between domain walls, equivalent to Ising spin configurations up to 
a global spin flip $s_i\rightarrow -s_i$, and dimer matchings on $G_D^*$.
For each set of domain walls, there is at least one corresponding dimer covering on the decorated dual lattice $G_D^*$.

The computation of the partition function $Z$ for the Ising model, a sum over all assignments of Ising spins,
can be directly expressed as a related sum over complete coverings of either $G^*$, the decoration of the
graph $G$ by Kasteleyn cities \cite{KasteleynIsing}, or coverings of $G_D^*$.
For sampling and computing
correlation functions, though, it is simpler to start with the decorated dual
graph \cite{ThomasMiddletonKastCities,ThomasMiddletonSampling}.
In a pure Ising model, summing over matchings on $G^*$ corresponds
to a high temperature expansion \cite{KacWard}, while sums over $G^*_D$ correspond to a low temperature expansion.
There is a simple correspondence between domain wall loops in $G_D^*$ and spin configurations:
given a set of domain walls, spins are found by setting spins within a single connected region to the same value.

The configurations contributing to the partition function sum correspond to terms in the expansion of the Pfaffian
of the Kasteleyn matrix \cite{KasteleynDimers}: to 
describe this correspondence, we first need to define the Kasteleyn matrix and the Pfaffian sum.
The Kasteleyn matrix $K$ is a skew-symmetric matrix with non-zero entries for each edge of the
decorated graph $G^*_D$; the rows and columns of the matrix are indexed by the vertices of the decorated graph.
Skew symmetry implies $K_{ab}=-K_{ba}$.
A non-zero entry $K_{ab}$ corresponds to an edge in $G^*_D$ connecting vertices $a$ and $b$. It is a matrix
of size $4N\times 4N$.
The values of the matrix are $\pm 1$ for edges internal to Kasteleyn cities. Edges that connect
cities have weights with absolute value $|K_{ab}|=\exp(-2\beta J_{ij})$,
where spins $i$ and $j$ have coupling $J_{ij}$ and the edge $ij$ in $G$ crosses
the edge $ab$ in $G^*_D$. The sign of each $K_{ab}$ is determined by a Pfaffian
orientation of the dimer graph (see, e.g., Ref.~\onlinecite{KrauthBook}).
For the graph $G^*_D$, a simple Pfaffian orientation is that horizontal edges between Kasteleyn cities
are oriented from left to right and vertical intercity edges are oriented from bottom to top,
so that if $a$ is to the left of $b$ or $b$ is above $a$, $K_{ab}>0$. The orientation of edges internal to a Kasteleyn city
can then be set as in Ref.~\cite{KasteleynIsing} or as described in Sec.~\ref{fig:MWAB}. Given a proper choice of signs for $K_{ab}$,
the partitition function for the original spin problem on a planar graph (without periodic boundaries) can then be shown
\cite{KasteleynDimers,Robertson} to be equal to
$\Pf(K)$, 
\begin{eqnarray}
Z=\sum_S e^{-\beta \mathcal{H}(S)} = \Pf(K),
\end{eqnarray}
where the Pfaffian of $K$ is defined by a sum over permutations $P$ of node indices,
\begin{eqnarray}\label{eq:PfK}
\Pf(K) = \sum_P \epsilon(P) K_{k_1l_1}K_{k_2l_2}\cdots K_{k_m l_m},
\end{eqnarray}
with $\epsilon(P)$ giving the sign of the permutation $P=(k_1,l_1,...,k_m,l_m)$
of the $M$ indices for the nodes of $G^*_D$ with $m=M/2=2L_xL_y$,
and the sum is restricted to the permutations satisfying the orderings $k_1<k_2<\ldots<k_m$ and 
$k_1<l_1$, $k_2<l_2$, ..., $k_m<l_m$. This choice of signs
forces all domain walls relative to a reference configuration to enter with a
positive sign; it also leads to the cancellation of terms such that the
many-to-one correspondence between dimer coverings and spin configurations
becomes one-to-one \cite{footnote_a}.
The permutation $P$ of indices that enters into the sum represents ``matchings'' or ``dimer coverings'', i.e., choices of
edges $e$, $e_1=(k_1,l_1)$, $\ldots$, $e_m=(k_m,l_m)$, such that each node in the dual decorated lattice belongs to exactly
one edge.
For proofs of the correctness of this mapping see, for example, Kasteleyn's papers \cite{KasteleynIsing} and
textbook treatments \cite{Robertson,KrauthBook}.
Note that the partition function for a graph of high genus (e.g., the three-dimensional Ising model) is impractical to compute,
as it requires a sum over a number of Pfaffians that is exponential in the genus \cite{Barahona,GLV}.

The computation of the partition function on planar graphs is simply given by the evaluation of a single Pfaffian.
Computations on a periodic graph are more complicated.
Kasteleyn described \cite{KasteleynDimers} how to compute the partition functions for dimers on a periodic, i.e.,
toroidal, lattice.
Four Pfaffians are computed for four variations of $K$, 
namely $K^{++}$, $K^{-+}$, $K^{+-}$, and $K^{--}$. We will refer to the set of these four matrices by the notation
$K^{\pm\pm}=\{K^{++},K^{-+},K^{+-},K^{--}\}$. For a given choice for $r\in\{+,-\}$ and $s\in\{+,-\}$, the matrix $K^{rs}$
has matrix elements $K_{ab}$ defined according to the standard Pfaffian orientation, except for those elements which
have endpoints $(a,b)$ at opposite ends of the square array: these elements correspond to the intercity edges
that wrap around the graph, leading to a periodic topology.
In the matrix $K^{rs}$, if an edge connects the a node $a$ in a city that is in column $L_x-1$ to a
node $b$ in column $0$,
its sign is given by $r$, while if an edge connects a node in a
city in row $L_y-1$ to a node for a city in row $0$, it has sign $s$.
These matrices $K^{\pm\pm}$ can be used to compute the partition functions $Z^{\alpha\beta}$ for $\alpha=\mathrm{P,AP}$
and $\beta=\mathrm{P,AP}$, where $\mathrm{P}$ indicates periodic boundary conditions along an axis and
$\mathrm{AP}$ indicates antiperiodic boundary conditions [negation of the $J_{ij}$ for all horizontal (vertical) edges in
a vertical (horizontal) line].
In particular, the partition functions
are given by linear combinations
\begin{eqnarray}
Z^{\alpha\beta}=\sum_{(rs)\in(\pm\pm)}L_{rs}^{\alpha\beta}\Pf\left(K^{rs}\right) \label{eqn:zab}
\end{eqnarray}
for a $4\times4$ matrix $L$ \cite{KasteleynDimers,ThomasMiddletonSampling},
\begin{eqnarray}\label{eq:L}
L=\frac{1}{2}\left\{\begin{array}{rrrr}
1&1&1&1\\
-1&-1&1&1\\
-1&1&-1&1\\
-1&1&1&-1
\end{array}\right\}\,.
\end{eqnarray}

Though the Pfaffian is formally written as a sum over a number of permutations that has
a number of terms roughly exponential in $N$, the Pfaffian of a general $N\times N$ matrix can be evaluated
in time polynomial in the number of nodes, in a fashion similar to computing the determinant.
However, given the two-dimensional nature of the graph underlying the matrix $K$, Pfaffians
(or determinants) and correlation functions can be computed
much more quickly (a lower power of $N$) than for a general matrix by splitting the
set of nodes geometrically in a hierarchical manner \cite{LRT}.

\section{Cluster matrices and their operations}\label{sec:main}

In this section, we first give an introductory outline to the  numerical methods we have implemented for
rapidly evaluating the partition function and correlation functions.
The details are then described in the subsections Sec.~\ref{subsec:PE} through Sec.~\ref{subsec:down}.
The algorithm for sampling configurations is described in Sec.~\ref{sec:sampling}.

The introduction to these methods requires the definition of the intermediate mathematical objects used,
the core mathematical steps applied to these objects, and the overall organization of these steps
to find the Pfaffian for the whole sample (or ratios of Pfaffians for correlation functions).

The Pfaffian for the whole sample is computed by combining information from smaller regions.
We can select a region $A$ on the decorated dual lattice $G_D^*$ by choosing a loop of Ising spins
on the spin lattice $G$: those nodes in $G_D^*$ that are ``inside''
the loop (generally the smaller set of nodes) will compose the interior set $A$ while those outside the loop
compose the exterior, complementary, set $\overline{A}$.
These geometrical regions or clusters have associated matrices and factors.
The central mathematical objects used in this procedure are antisymmetric ``cluster'' matrices $U_A(\mathcal{J})$
and $U_{\overline{A}}(\mathcal{J})$ \cite{LRT,FIND} which depend both on the set of spin couplings $\mathcal{J}=\{J_{ij}\}$
and the region $A$.
The dependence of $U$ on the spin couplings $\mathcal{J}$ that define the given realization of a sample will be implicit
in the remainder of this paper and so we will write $U_A$ for $U_A(\mathcal{J})$.
A given cluster matrix is indexed by the nodes
of the decorated dual lattice that are on the boundary of the clusters: if there are $m$ boundary vertices in the cluster,
the matrix $U$ has dimensions $m\times m$. The boundary correlations of dimers (and hence
spins on the original graph) are directly related to the cluster matrices $U$ by a matrix inverse.
Also associated with each region $A$ is a factor $z(A)$, the ``partial Pfaffian''.
This factor represents a multiplicative contribution to the overall
partition function.
It represents a sum over dimer configurations on the interior of $A$.

The core mathematical steps applied to the cluster matrices are the collection of cluster matrices
for neighboring regions into a larger matrix and subsequent elimination (contraction) steps applied
to this joint matrix.
These elimination steps remove rows and columns from the joint matrix that correspond to
nodes that are on the boundary of the original neighboring regions but are not boundary nodes for the
union of the two regions.
The remaining matrix is then indexed by the boundary nodes of the larger, unified region.
This removal of nodes is carried out by Pfaffian elimination,
a procedure described in Sec.~\ref{subsec:PE} and one that is similar
to Gaussian elimination. This directly implements a sum over the
the dimer coverings over edges that are shared by the adjacent clusters and incorporates that sum into the
partial Pfaffian factor.
To collect neighboring regions $A$, with $m_A$ boundary nodes, and $B$, with $m_B$ boundary nodes,
a square matrix of size $(m_A+m_B)\times(m_A+m_B)$ is filled with the elements of $U_A$ and $U_B$, in
block diagonal form,\begin{eqnarray}
M_0(W,A,B)=\left(\begin{array}{cc}U_A&W_{ab}\\-W_{ab}^T&U_B\end{array}\right)\ .\label{eq:fill}
\end{eqnarray}
where the matrix $W_{ab}$ is indexed by the boundary nodes of $A$ and $B$ and has nonzero elements when $a$ and $b$ are the
ends of an intercity edge $e_{ab}$ connecting $A$ to $B$.
Partial Pfaffian elimination then removes from the matrix rows and columns
that correspond to the nodes belonging to separating edges in $W$, while maintaining the overall Pfaffian.
The matrix resulting from elimination
will be the cluster matrix $U_C$ for the joined regions $C=A\cup B$, with the matrix again indexed by the remaining
boundary nodes. The matrix $U_C$ has dimension $m_C=m_A+m_B-2|W|$.
The two steps together, collection and Pfaffian elimination, will be referred to
as a ``merger''.

The methods for evaluating the partition function $Z=\Pf(K)$ are based on relating the Pfaffian of a region of the sample
to the Pfaffians defined for subregions: by recursive application of this relationship, the Pfaffian $\Pf(K)$ of the whole sample
can be computed. At the largest scale of this recursion, for example, it turns out that we can write
\begin{eqnarray}
\Pf(K) = \sigma(A^*,B^*) z(A^*)z(B^*)\Pi_{e\in W} x_e \label{finalanswer}
\end{eqnarray}
where the sample is divided geometrically into two regions, $A^*$ and $B^*$, the $z(A)$ and $z(B)$ factors
are ``partial Pfaffians'' computed recursively. The factors $x_e$ result from the Pfaffian elimination steps
described in Sec.~\ref{subsec:PE}. The prefactor $\sigma(A^*,B^*)=\pm 1$ is determined by the sign of how
the rows and columns of $U_{A^*}$ and $U_{B^*}$ are combined. In turn,
we can write, for example,
\begin{eqnarray}
z(A^*)=\sigma(A_1,A_2)z(A_1)z(A_2)\Pi_{e\in W_A}\,x_e\label{eq:zsigmaPi}
\end{eqnarray}
where the region $A^*$ is decomposed into regions $A_1$ and $A_2$ and $W_A$ is the set of edges that connect these two
sets of nodes.
Note that the parity factors $\sigma$ are not strictly needed for computing the partition function in planar graphs, as all that matters
in that case is the magnitude of $\Pf(K)$, but they are needed whenever periodic boundary conditions are used. For numerical
stability, pivoting operations that permute the rows and columns are used and the choice of
pivots may be different for the distinct $K^{\pm\pm}$.

The organization of the cluster matrix mergers is divided into two stages, the up sweep stage and the down sweep stage \cite{FIND}.
In each sweep, matrices representing
neighboring or enclosing regions are merged. The organization of these mergers
is set by the recursive geometric division of the sample. In the up sweep stage, smaller
cluster matrices $U_A$ and $U_{B}$ for neighboring clusters $A$ and $B$ are merged to create a cluster matrix $U_C$ for
the union $C=A\cup B$ of the two clusters. This information sums information over smaller scales into information at larger
scales. The up sweep stage is sufficient to compute the partition function of a sample.
In the down sweep stage, correlations (and configuration samplings) can be computed.
In this stage, the sum of statistical weights of all dimer configurations external to a region is used to find the
sum of statistical weights external to smaller regions. If $C$ is the
union of clusters $A$ and $B$, $U_{\overline{C}}$ gives the matrix encoding the sum of statistical
weights external to the region $C$.
This matrix is originally found by summing over all
dimer configurations external to the region $C$. The cluster matrix $U_{\overline{C}}$
can be merged with $U_B$.
This sums over the configuration sums internal to $B$
and the dimer configurations external to both $A$ and $B$, giving a matrix defined on the boundary of $A$
that represents the sum over dimer configurations external to $A$,
the cluster matrix $U_{\overline{A}}$. At each stage of this recursion,
the cluster matrices $U_A$ and $U_{\overline{A}}$ can then be used together to find correlations on the boundary of $A$.
It turns out that the sums of signed mergers of these two matrices
gives the spin-spin correlation functions for the Ising spins that lie between $A$ and $\overline{A}$.

For reference and to provide a flavor of the methods, we present an outline of the steps for computing the partition function and
correlation functions; more detailed descriptions of these steps are given in the subsequent subsections:
\begin{enumerate}
\item{From the bond weights $J_{ij}$, generate the weights $w_{ij}=e^{-2\beta J}$ for all neighboring spins in the lattice.}
\item{Generate a binary tree $T$ for the geometric
subdivision of the decorated dual lattice $G_D^*$. Each node of the tree contains geometric information for a region $A$,
the cluster matrices $U_A$ and $U_{\overline{A}}$, and the partial Pfaffian factors $z(A)$.
All non-leaf nodes of the tree have pointers to two children representing matrices for two subregions of approximately the same size.
The subdivision is terminated at the scale of Kasteleyn cities, which are regions that correspond to the leaves of $T$.}
\item{Up sweep: starting from the leaves of $T$, 
merge sibling pairs of cluster matrices $(U_A,U_B)$ and factors $z(A)$ and $z(B)$
to compute parent matrices $U_C$ and partial Pfaffian factors $z(C)$.}
\begin{enumerate}
\item{This merging is initiated by collecting the matrices $U_A$ and $U_B$ along with
edge weights for the edges $W$ connecting $A$ and $B$ together into a joint matrix $M_0(W,U_A,U_B)$
(see \myeqref{eq:fill}).}
\item{Pfaffian elimination then reduces the matrix $M_0$ into a set of factors $x_e$ and a smaller
matrix $U_C$ indexed by the boundary of $A\cup B$.}
\item{Set $z(C)=\sigma(A,B) z(A)z(B)\Pi_{e\in W}x_e$, where $\sigma(A,B)=\pm 1$ gives the total parity of row/column
permutations that were used in the rearrangements of $M_0$ in preparation for Pfaffian elimination and the parity of permutations
used for pivoting steps during the Pfaffian elimination.}
\item{
These up sweep steps are carried out recursively, merging clusters
up to, but not including, the last pair $A^*$ and $B^*$ representing the initial division of the whole sample.}
\end{enumerate}
\item{The two largest clusters for $A^{*}$ and $B^{*}$ are then merged according to the choice of boundary conditions:}
\begin{enumerate}
\item{For open or fixed boundary conditions, simply merge the two top-level cluster matrices $U_{A^*}$ and $U_{B^*}$.
In this case, Pfaffian elimination eliminates all rows and columns and the partition function $\Pf(K)$ is given by \myeqref{finalanswer}.}
\item{For periodic boundaries, merge the $U_{A^*}$ and $U_{B^*}$ along one of the rows or columns separating them (there are either two rows or two columns separating them for periodic BCs) into a matrix $U_{C^*}$. Then connect the matrix $U_{C^*}$
with itself along a remaining row to generate two matrices $U_{C^*}^+$ and $U_{C^*}^-$, the former using positive weights for the wrapping edges, the latter using negative weights. Eliminate those connecting edges. Then
include wrapping edges along the remaining axis, again using negative and positive edge weights for each of the $U_{C^*}^\pm$.
The resulting eliminations give scalars: these overall weights are the Pfaffians $\Pf(K^{\pm\pm})$.}
\item{Compute $Z^{\mathrm{P,P}}$, $Z^{\mathrm{AP,P}}$, $Z^{\mathrm{P,AP}}$, $Z^{\mathrm{AP,AP}}$ from linear
combinations of $\Pf(K^{\pm\pm})$, as given by \myeqref{eqn:zab}.}
\end{enumerate}
\item{Stop here if only the partition function is required. Continue to the next steps to compute correlation functions.}
\item{Down sweep: descend the tree $T$, computing cluster matrices for complementary regions and merging interior
and exterior matrices to find correlation functions:}
\begin{enumerate}
\item{Use the results of the up sweep to initialize the two top level complementary
cluster matrices via $U_{\overline{A^*}}=U_{B^*}$ and $U_{\overline{B^*}}=U_{A^*}$.}
\item{If periodic boundary conditions are used, merge $U_{A^*}$ and $U_{\overline{A^*}}$ and merge
$U_{B^*}$ with $U_{\overline{B^*}}$ using the four different choices for wrapping edge weights, i.e., select
$(r,s)$ from $(\pm,\pm)$. Use these mergers to set up four parallel trees for further descent.}
\item{For all down sweep steps for a planar Ising model or
further descending steps in the case of periodic boundary conditions in each of the four trees:}
\begin{enumerate}
\item{Given a parent $C$ with known $U_{\overline{C}}$ and children $A$ and $B$,
merge the parent matrix $U_{\overline{C}}$ with $U_B$ to generate matrices $U_{\overline{A}}$ for regions complementary to $\overline{A}$,
as in the FIND algorithm \cite{FIND}.}
\item{Also merge $U_{\overline{C}}$ with $U_A$ to generate $U_{\overline{B}}$.}
\end{enumerate}
\item{Compute correlation functions between spins on the corners of any given region $A$ by signed merging of $U_{\overline{A}}$ and
$U_A$. (To find correlation functions for periodic boundary conditions, compute the correlation function as the weighted
sum over four trees as given by \myeqref{weightedCorrs}.)}
\end{enumerate}
\end{enumerate}

\subsection{Pfaffian elimination}\label{subsec:PE}
Pfaffian elimination simplifies a matrix by setting chosen elements in a row to zero while maintaining
the Pfaffian of the matrix as an invariant. This elimination proceeds by a process similar
to Gaussian elimination for general matrices, but is applied to skew-symmetric matrices \cite{Bunch}.
In Gaussian elimination, the lower triangular elements are set to zero and the determinant is the product
of the diagonal elements.
In Pfaffian elimination,
the diagonal elements of a given skew-symmetric $U$ are zero and Pfaffian elimination aims to set
all elements that are more than one step off of the diagonal to zero. The Pfaffian of the matrix is the product of
the remaining elements in even-indexed rows (given that the first row has index 0).

Pfaffian elimination can be defined inductively for a skew-symmetric matrix $U$. Each step simplifies one row
to a single non-zero element.
Suppose that Pfaffian elimination has been carried out for rows with index less than $i$, where $i$ is even
and the rows are indexed starting with row $0$ and that the element in row $i$ and column $i+1$ is non-zero.
Then multiples of row $i$ and column $i+1$ can be added to rows and columns of higher index to
zero out the remaining elements of row $i$ and column $i$.
This addition of rows and columns simplifies the matrix while the Pffafian is unchanged, in the same fashion as row and column
additions in a matrix do not modify its determinant.
Specifically, for $j>i+1$, column $i+1$ is multiplied by $-U_{i,j}/U_{i,i+1}$  and added to column $j+1$
and row $i+1$ is multiplied by the same prefactor and added to row $j+1$ \cite{Bunch}.
Note that the odd rows do not contribute
to the Pfaffian when the elimination in the previous even row is completed, so that elimination is
applied only to even rows.
We use pivoting of the rows and columns that are to be eliminated
to improve numerical stability. A pivot is an interchange between indices $c$ and $d$: the elements of row $c$
are swapped with the elements of row $d$ at the same time columns $c$ and $d$ are swapped.
As we use it here, Pfaffian elimination is often carried out only for some subset of rows.
Note that rows/columns that are not to be eliminated are not considered
for pivoting. The permutations due to pivoting operations place the element
with the largest available magnitude in the superdiagonal position, before the elimination is carried out.
Each pivot leads to a change of sign in $\Pf(U)$ which is accumulated in the prefactor $\sigma$.

Mathematically, Pfaffian elimination carried out for all rows can be used as a factorization scheme, 
similar to LU factorization via Gaussian elimination \cite{Bunch}.
The Pfaffian elimination procedure applies linear operations to $U$ so that $LUL^T=F$
where $L$ is a lower triangular matrix
and $F$ is zero except for the superdiagonal elements. The inverse of a skew-symmetric matrix $U$ is then
\begin{eqnarray}
U^{-1}=L^TF^{-1}L\,;\label{eq:PfInverse}
\end{eqnarray}
this procedure of elimination and matrix multiplication is used to find matrix inverses
in the sampling of Ising spin configurations (see Sec.~\ref{sec:sampling} and Ref.~\cite{ThomasMiddletonSampling}).
In the mergings of matrices used here, the rows and columns corresponding
to nodes on the boundary of the joined regions are kept, while the rows and columns corresponding
to nodes shared by the joined regions are eliminated.
The eliminated rows and columns have superdiagonal elements which are multiplied together to give
a partial Pfaffian while the rows and columns for the new boundary are carried onto the next stage.
Physically, by eliminating rows and columns corresponding to nodes internal to a geometric region, these
steps ``integrate out'' degrees of freedom internal to the new cluster.

\subsection{Geometric dissection}\label{subsec:geom}
Computations for sparse matrices that are derived from two-dimensional graphs
can be very efficiently carried out using the important technique of nested dissection \cite{LRT}.
The row and column indices of the matrix correspond to a numbering of the nodes in a graph.
The idea behind nested dissection is to hierarchically subdivide the matrix according to
row and column indices that index nodes for distinct compact regions.
When subdividing a region into two child regions,
the separator for this subdivision can be taken to be either nodes that lie between the two regions or
a set of edges that connects the two compact subdivisions.
By``compact'', we mean regions of size $N$ whose boundary 
scales as $O(\sqrt{N})$. The result for the parent region is found
by separately computing the results for the two child regions and stitching those two results together
using the separator. As a separator can be found with $O(\sqrt{N})$ nodes for a matrix
of scale $N\times N$ (i.e., of order $L$ for a spatial region of size $L^2=N$),
with the two subproblems of comparable size, the computation at each
scale is for matrices of size $O(\sqrt{N})\times O(\sqrt{N})$ \cite{LRT}. The work at each scale
is therefore much less than for the dense case where the problem cannot be efficiently
subdivided and one needs to consider matrices of size $N\times N$.
The first application of nested dissection to
efficiently computing spin glass partition functions is described in Ref. \cite{GLV}. 
The use of the general concept of nested dissection for
sampling dimer configurations was proposed in Ref.  \cite{Wilson} and carried out
for Ising spin glasses in Ref. \cite{ThomasMiddletonSampling}.

In the form of nested dissection \cite{LRT} used for dimer sampling \cite{Wilson},
a set of nodes in a graph is selected as the separator.
This is the form we previously used \cite{ThomasMiddletonSampling} for sampling Ising spin
configurations.
Here, we instead use an edge separator with each separating edge having one node in each child region.
An example approach that inspired our method is the FIND (fast inverse using
nested dissection) technique, which computes some of the elements of an inverse
matrix, as used in computing non-equilibrium Green's functions in a two-dimensional quantum device. The
asymptotic run-time of computing the Pfaffian with either node or edge separators scales with $N$ in
the same way, i.e., as $O(N^{3/2})$ but the FIND approach has several advantages for studying the
Ising model.
These advantages include simplifying the structure of the code as well as allowing for more
direct computations of the inverse matrix elements and the Pfaffian ratios used to
sample configurations.

\begin{figure}[h]
\centering
(a)\includegraphics[width=1.4in]{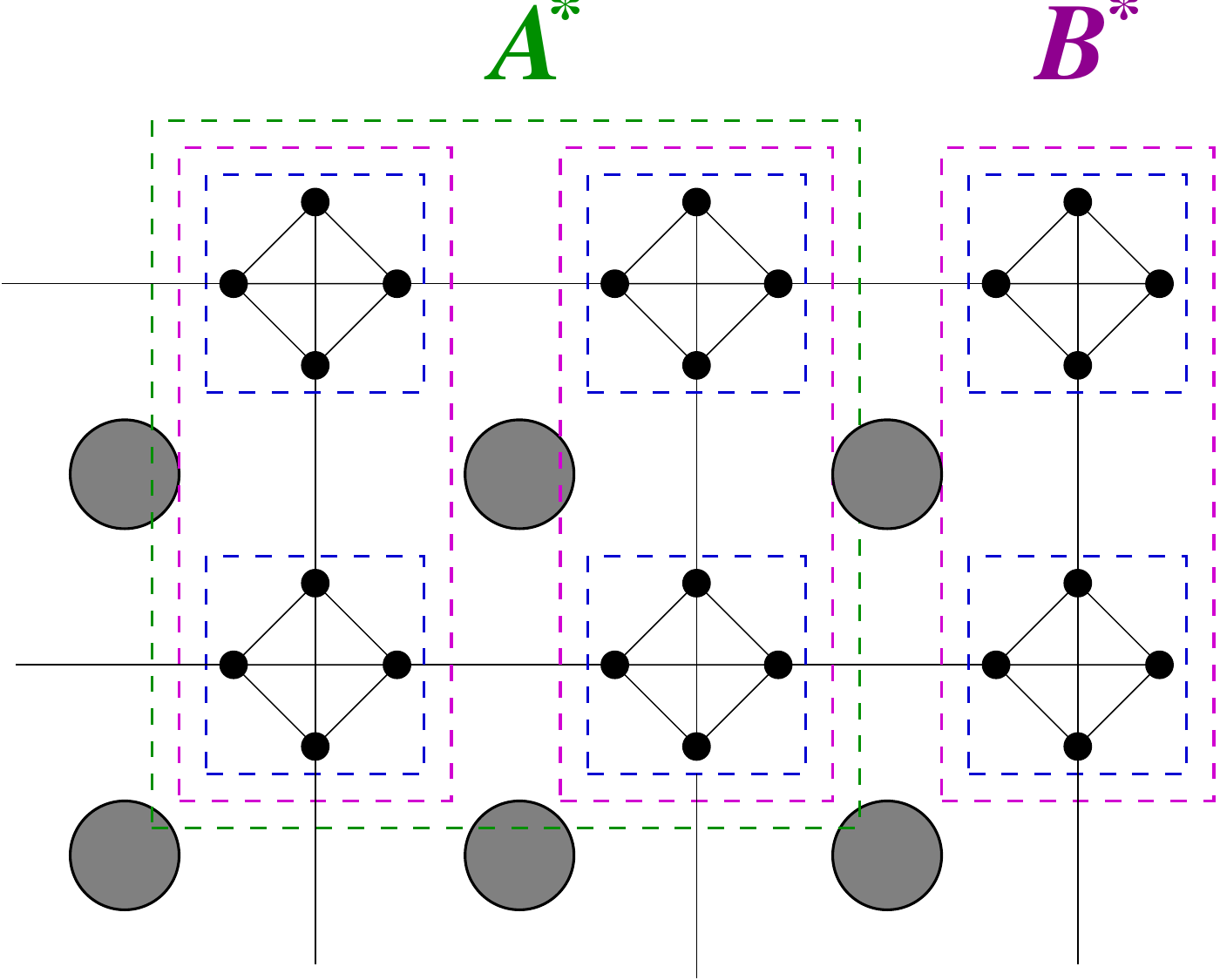}
\ \ (b)\ \ \includegraphics[width=1.2in]{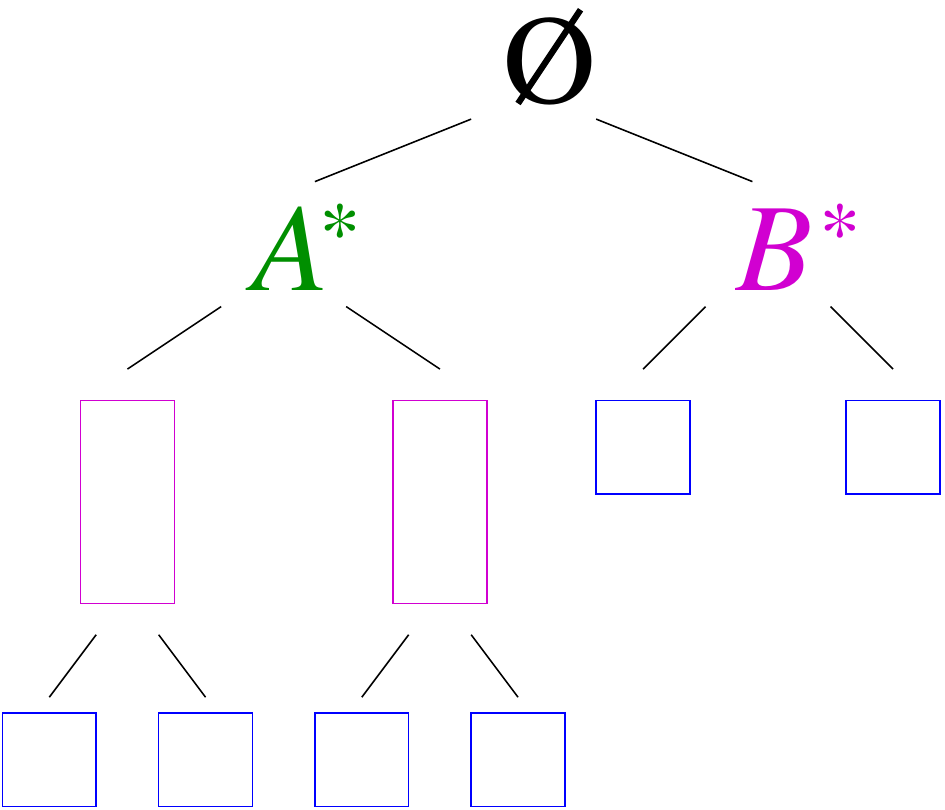}
\caption{(color online) Depictions of the geometric dissection tree $T$ for an example decorated dual graph. The original
Ising system has $L_x\times L_y = 3\times 2$ spins on a periodic graph; the spins are indicated by gray circles. The decorated dual
graph has $3\times 2$ Kasteleyn cities (the sets of four fully connected nodes). At each stage,
a parent rectangle of Kasteleyn cities is divided into two roughly equal sibling rectangles, the children.
In the algorithms described here, each rectangle $A$ has geometric information, two cluster matrices
$U_A$ and $U_{\overline{A}}$,
and a partial Pfaffian factor $z(A)$.
(a) The nested dissection in real space. The two largest subregions with boundaries, $A^{*}$ and $B^{*}$, indicated.
The region $C^*$ in the final stage (not shown for clarity) is the union of $A^{*}$ and $B^{*}$, $C^{*}=A^*\cup B^*$;
at the end of the algorithm, it has no boundary
but a matrix corresponding to this region is used as a working matrix when using periodic boundaries.
(b) A diagram of the resulting nested dissection tree $T$. The leaves of the tree are Kasteleyn cities.
The root of the tree has no cluster matrix associated with it, as the sample has no boundary, though it has a Pfaffian
factor associated with it, which is used to find the partition function for the whole sample.
}
\label{fig:tree}
\end{figure}

In the Ising model, the decorated dual graph $G_D^*$ for an $L_x \times L_y$ square sample with periodic boundaries can
be recursively divided by splitting it either horizontally or vertically at each stage into
smaller rectangles. \figref{fig:tree} gives an example of this dissection. The geometric dissection of the system into smaller
rectangles is described by a binary tree $T$.
Each rectangle is an array of Kasteleyn cities. The leaves of the tree consist of $1\times 1$ arrays, that is, individual
Kasteleyn cities, so that the corresponding cluster matrix $U_Y$ for a city $Y$ is a $4\times 4$ matrix.
At each stage of the dissection,
the graph is divided along the axis with the shortest length and as close to the middle of the rectangle
as possible. This division splits the city set by cutting the edges which join neighboring cities; these
cut edges comprise the separating set $W$ at each stage.
It is important to note that the separator $W$ has a corresponding set of Ising spins:
the spins that lie between the two geometric regions and are separated from each other by the edges
in the set $W$ (see \figref{fig:sep}).
The top of the tree $T$ has no boundary and so is associated with a null matrix at the end of the algorithm.
However, for efficiency in collecting partial results, the region $C^*$ corresponding to the whole sample
has matrices associated with it during intermediate stages of the calculation.
The two rectangles $A^{*}$ and
$B^{*}$ that result from the first division of the sample are the first non-empty regions, with
$A^*\cup B^*=C^*$.

This tree structure is used to organize the elimination steps in the FIND-based technique,
which consists of two stages:
an up sweep which produces the partition function of the system by Pfaffian elimination,
and a down sweep which may be used to find inverse matrix elements, bond probabilities, or
correlation functions.  These stages may be understood as a reorganization to move 
information about dimer correlations on the region boundaries from one
scale to another.  First, in the up sweep stage, the cluster matrices, which represent boundary
information about couplings that remains after summing over internal degrees of freedom,
are joined with cluster matrices in neighboring regions to generate cluster information at a
larger scale, for the joint region. This is repeated until the aggregate thermodynamic properties of
the entire sample are found. Next, in the down sweep stage,
this information may be propagated back down to give correlation results at
the smallest scales, and all scales in between. This propagation is effected
by merging the correlation information exterior
to a region with the interior correlation information.

\subsection{Up sweep stage}\label{subsec:up}
The matrix operations for the Pfaffian eliminations carried out in the up sweep stage can be illustrated by an example
of the first steps of the algorithm. Describing these first steps allows us to display the matrices used and their
correspondence to the graphs showing dimer correlations.
Subsequent steps use larger matrices, but have the same structure.

The lowest level steps merge the cluster matrices
for two neighboring Kasteleyn cities. Let two such cities be denoted by $A$ and $B$.
These cities each correspond to neighboring nodes on the dual square lattice $G_D$ and 
are two of the leaves of the tree representing the geometric dissection of $G_D^*$.
The corresponding Kasteleyn matrices, $U_A$ and $U_B$, are the simplest cluster matrices.
The rows and columns of $A$ and $B$ are each indexed by four nodes,
so $U_A$ and $U_B$ are each of size $4 \times 4$.
In general, due to skew symmetry, only the upper triangular portion of each cluster matrix need be stored in memory.
The elements above the diagonal in the matrices $U_A$ and $U_B$ are displayed in \figref{fig:MWAB}(a).
Each cluster matrix has a weight associated with it, a partial Pfaffian that accumulates the weights of eliminated rows,
which is initialized to be unity, $z(A)=z(B)=1$.
The entries of each matrix have weight of magnitude $1$, with signs appropriate for Kasteleyn cities \cite{KasteleynIsing}.
For the sign conventions and numbering scheme show in \figref{fig:MWAB}(a), all elements of $U_A$ and $U_B$ are
positive in the upper triangular section.
The edges joining cities are directed in the positive $x$ and positive $y$ directions,
so that the matrix elements $K^{++}_{kl}$ are non-negative for nodes $k$ in cities to the
left of or below the city containing node $l$.
Here, the separator $W$ consists of a single edge $e$ joining city $A$ to city $B$.
The weight of this edge is $w=\exp(-2\beta J_{ij})$, where $J_{ij}$ is
the bond weight on the connection between spins $i$ and $j$ that is perpendicular to this dual edge $e$.
Note that this edge is not connected to the rest of the lattice and so will be
be eliminated when merging $A$ and $B$.
To carry out this reduction, the elements of $U_A$ and $U_B$ and the edge weight are
copied into a joint temporary matrix $M_0(W,U_A,U_B)$, which is of size $8\times 8$ (28 upper triangular elements).
This edge to be eliminated is then
placed in the first row of the matrix by permuting the rows of the joint matrix to obtain $M_1(W,U_A,U_B)$.
Whenever two rows are interchanged, an overall minus sign is introduced into the Pfaffian factors.
In this simple case, only one Pfaffian elimination is applied to $M_1(\{w\},U_A,U_B)$. This eliminates the
connections of the ends of the connecting edge to the rest of the boundaries of $A$ and $B$, giving a matrix
with the first superdiagonal element $x_{1,2}$ is non-zero, but the rest of the first row eliminated.
The remaining rows, the third through the last rows, define the new cluster matrix $U_C(W,U_A,U_B)$.
This matrix
encodes the correlations along the outer boundary of $C$, the region composed of  the two joined cities.
This contracted matrix is generally not sparse; see \figref{fig:MWAB}.
Using \myeqref{eq:zsigmaPi},
the partial Pfaffian factor that is stored along with $U_C$ is $z(C)=-z(A)z(B)x$, where the minus sign is
included because of the row interchange.

\begin{figure}[h]
\centering
\includegraphics[width=3.0in]{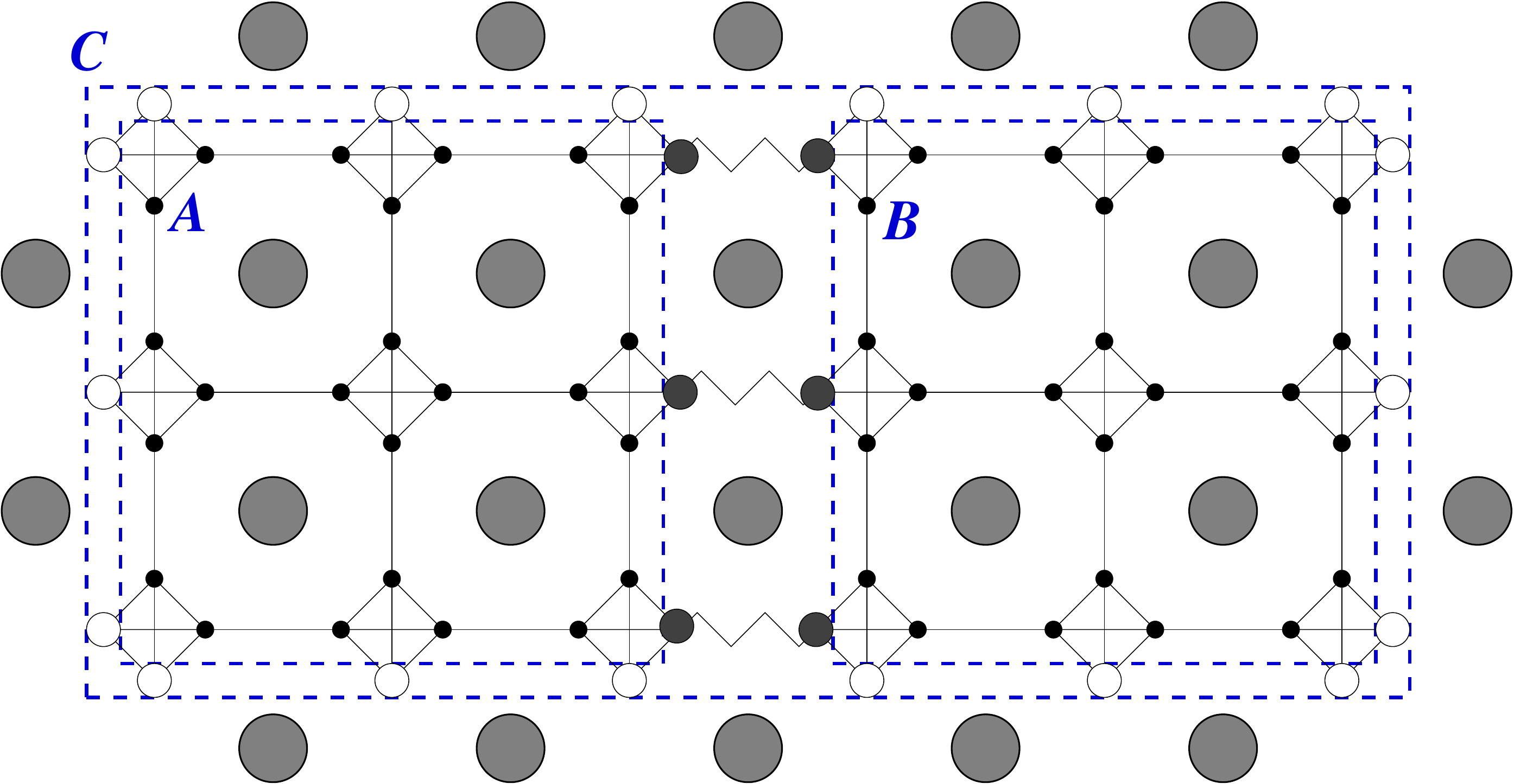}
\caption{Example of a higher level step of the up sweep stage.
The Ising spins on the original lattice whose couplings $J_{ij}$ are relevant to the calculations through this
step are indicated by the large grey circles, while the
nodes of the decorated dual lattice are indicated by the medium-sized and small circles. The solid straight
and jagged lines indicate edges belonging to the decorated dual graph. Two geometric
regions on the dual lattice, $A$ and $B$, each containing $3\times 3$ Kasteleyn cities, are denoted by the (blue) dashed squares.
The dual edges $W$ that separate $A$ and $B$ are drawn as jagged lines and join the filled in medium-sized nodes.
The small nodes are interior to the regions $A$ and
$B$ while the larger nodes (medium-sized open and filled circles) form the borders of $A$ and $B$.
In the up sweep stage (see Sec.\ \ref{subsec:up}), given the cluster matrices $U_A$ and $U_B$ for $A$ and $B$,
the separating edges belonging to the separator $W$ are integrated out via Pfaffian
elimination, leaving a cluster matrix for $C$ that includes the entire subgraph shown.
The cluster matrix $U_C$ is indexed by its border nodes, i.e., the medium-sized open circles connected by the dashed (blue) rectangle.}
\label{fig:sep}
\end{figure}

This process of merging adjacent subgraphs of $G^*_D$, which uses
Pfaffian elimination to remove adjacent boundary nodes, is
repeated at each scale up to the system size $L$.  Generally, neighboring regions $A$ and $B$ are
joined together by copying their entries into a joint matrix $M_0$, adding the weights of connections
for the set of $n$ edges $W$ that join $A$ to $B$ ($W$ is indicated by
jagged lines in \figref{fig:sep}), permuting the joint matrix  to give $M_1$,
and then eliminating the first $|W|$ rows. 
The portion of the matrix that is indexed by the boundary of $C=A\cup B$ is the larger scale cluster matrix $U_C$.
The product of the superdiagonals on the even eliminated 
rows are used to find $z(C)=\sigma(P)z(A)z(B)\Pi_{i=0}^{2(|W|-1)}x_{i,i+1}$, where $\sigma(P)$ is the sign of the permutations
carried out in assembling and carrying out pivot eliminations during the elimination and the $x_{i,i+1}$ for even $i$ for
the eliminated edges are the superdiagonal elements remaining after Pfaffian elimination.
This process is an exact real-space renormalization
process on the space of cluster matrices.  At each scale, the cluster matrices represent geometric regions
whose interactions are computed using their adjacent boundaries, though each of these clusters has many internal degrees of
freedom.  At the largest length scale, the Pfaffian of the remaining
$\mathcal{O}(L)\times\mathcal{O}(L)$ matrix is multiplied by
the products $z(A^*)$ and $z(B^*)$ resulting from all lower level mergers to gives the Pfaffian of the entire Kasteleyn matrix, i.e., the
partition function $\Pf(K)$.  This procedure of Pfaffian elimination and collection of superdiagonal elements preserves the overall
Pfaffian at each stage, since Pfaffian elimination maintains the Pfaffian as an invariant and the Pfaffian of a matrix with only superdiagonal
elements in the odd rows is just the product of those superdiagonal elements. The total number of operations in a full up sweep
is dominated by the last merger and is of order $\mathcal{O}(N^{3/2})$.

\begin{figure}[t!]
\centering
\setlength{\unitlength}{0.009cm}
\begin{picture}(660,375) 
\put(10,350){\makebox(0,0){$(a)$}}
\put(220,290){\line(1,1){60}}
\put(220,290){\line(1,-1){60}}
\put(280,230){\line(0,1){120}}
\put(220,290){\line(1,0){120}}
\put(340,290){\line(-1,1){60}}
\put(340,290){\line(-1,-1){60}}
\put(220,290){\circle*{15}}
\put(280,350){\circle*{15}}
\put(280,230){\circle*{15}}
\put(340,290){\circle*{15}}
\put(195,290){\makebox(0,0){$3$}}
\put(340,315){\makebox(0,0){$1$}}
\put(250,230){\makebox(0,0){$2$}}
\put(250,350){\makebox(0,0){$0$}}
\put(520,230){\line(0,1){120}}
\put(460,290){\line(1,0){120}}
\put(580,290){\line(-1,-1){60}}
\put(580,290){\line(-1,1){60}}
\put(460,290){\line(1,1){60}}
\put(460,290){\line(1,-1){60}}
\put(580,290){\circle*{15}}
\put(520,350){\circle*{15}}
\put(520,230){\circle*{15}}
\put(460,290){\circle*{15}}
\put(460,315){\makebox(0,0){$3$}}
\put(600,290){\makebox(0,0){$1$}}
\put(545,230){\makebox(0,0){$2$}}
\put(545,350){\makebox(0,0){$0$}}
\put(340,290){\line(1,0){120}}
\put(340,265){\makebox(110,20){$w=0.5$}}
\put(100,10){\makebox(200,160)
{ $U_A = \left(
  \begin{array}{rrr}
  1 & 1 & 1 \\
    & 1 & 1 \\
    &   & 1 \end{array} \right)$
  }
}
\put(500,10){\makebox(200,160)
{ $U_B = \left(
  \begin{array}{rrr}
  1 & 1 & 1 \\
    & 1 & 1 \\
    &   & 1 \end{array} \right)$
  }
}
\end{picture}

\begin{picture}(660,620) 
\put(10,580){\makebox(0,0){$(b)$}}
\put(24,340){\makebox(660,240)
{ $M_0 = \left(
  \begin{array}{rrrrrrrr}
  1 & 1 & 1 & 0 & 0 & 0 & 0 \\
    & 1 & 1 & 0 & 0 & 0 & 0.5\\
    &   & 1 & 0 & 0 & 0 & 0\\
    &   &   & 0 & 0 & 0 & 0 \\
    &   &   &   & 1 & 1 & 1 \\
    &   &   &   &   & 1 & 1 \\
    &   &   &   &   &   & 1 \\
  \end{array}
  \right) $ }
  }
\put(24,30){\makebox(660,240)
{ $M_1 = \left(
  \begin{array}{rrrrrrr}
  0.5 & -1 &  0 &  0 &  0 &  1 &  1 \\
      &  0 & -1 & -1 & -1 &  0 &  0 \\
      &    &  0 &  0 &  0 &  1 &  1 \\
      &    &    &  1 &  1 &  0 &  0 \\
      &    &    &    &  1 &  0 &  0 \\
      &    &    &    &    &  0 &  0 \\
      &    &    &    &    &    &  1 \\
  \end{array}
  \right) $ }
  }
\end{picture}

\begin{picture}(660,400) 
\put(10,385){\makebox(0,0){$(c)$}} 
\put(220,320){\line(1,1){60}}
\put(220,320){\line(1,-1){60}}
\put(280,260){\line(0,1){120}}
\put(220,320){\circle*{15}}
\put(280,380){\circle*{15}}
\put(280,260){\circle*{15}}
\put(195,320){\makebox(0,0){$5$}}
\put(250,260){\makebox(0,0){$4$}}
\put(250,380){\makebox(0,0){$0$}}
\put(520,260){\line(0,1){120}}
\put(580,320){\line(-1,-1){60}}
\put(580,320){\line(-1,1){60}}
\put(580,320){\circle*{15}}
\put(520,380){\circle*{15}}
\put(520,260){\circle*{15}}
\put(600,320){\makebox(0,0){$2$}}
\put(545,260){\makebox(0,0){$3$}}
\put(545,380){\makebox(0,0){$1$}}
\put(220,320){\line(1,0){360}}
\put(220,320){\line(5,1){300}}
\put(220,320){\line(5,-1){300}}
\put(280,380){\line(1,0){240}}
\put(280,260){\line(1,0){240}}
\put(280,380){\line(2,-1){240}}
\put(280,380){\line(5,-1){300}}
\put(280,260){\line(2,1){240}}
\put(280,260){\line(5,1){300}}
\put(36,0){\makebox(600,240)
{ $U_C = \left(
  \begin{array}{rrrrr}
  -2 & -2 & -2 &  1 &  1  \\
     & 1  &  1 & -2 & -2 \\
     &    &  1 & -2 & -2 \\
     &    &    & -2 & -2 \\
     &    &    &    &  1  \\
  \end{array}
  \right) ; z(C)=0.5$ }
  }
\end{picture}
\caption{Depiction of merger of Kasteleyn cities $A$ and $B$ connected by
an edge of weight $w=\exp(-2\beta J_{ij})$, with $w$ set to $0.5$. (a) The rows and columns of the $U_A$ and $U_B$
are indexed by nodes $\{0,1,2,3\}$ and the upper triangular part of these 
skew-symmetric matrices is shown with the diagonal of zeros
omitted. For example, the upper left elements shown in the matrices here are in row 0 and column 1.
The signs of the connections correspond to a Pfaffian orientation \cite{KasteleynIsing} to consistently count spin configurations.
(b) The matrix $M_0$ is the result of collecting $U_A$ and $U_B$ and the edge weight
$w=0.5$ and is indexed by nodes $0$ through $7$, with indices from $A$ for the initial part $\{0,1,2,3\}$ and 
indices from $B$ for the second half $\{4,5,6,7\}$.
Permuting rows/columns $0$ and $1$ and then $1$ and $7$ gives the matrix $M_1$.
These permutations place the nodes for the edge to be eliminated in the first two rows of $M_1$.
(c) After using Pfaffian elimination
to remove rows and columns $0$ and $1$, one is left with a superdiagonal element at $(0,1)$ of 0.5 (no pivoting is possible in this case)
giving a partial Pfaffian $z(C)=0.5$
and the next generation cluster matrix $U_C$, indexed by the remaining 6 nodes numbered as shown.
}
\label{fig:MWAB}
\end{figure}

\subsection{Down sweep stage: overview}\label{subsec:downoverview}
While the up sweep stage can be used to compute a global quantity, e.g., the partition
function for given boundary conditions, the subsequent down sweep stage provides
a powerful method for computing spatial information such as spin-spin correlation functions.
Correlation functions at multiple scales can be computed in a single down sweep, while
multiple down sweeps are used to generate sample configurations (see Sec.\ \ref{sec:sampling}),

The down sweep stage descends the geometry tree $T$, recursively computing new cluster matrices $U_{\overline{A}}$.
These matrices contain information about sums over dimer configurations for the exterior $\overline{A}$ of the geometric regions $A$.
These exterior clusters are merged with the interior cluster matrices that were computed on the up sweep to
find spin-spin correlation functions. The computed correlation functions, i.e., the thermal averages $\langle s_i s_j \rangle$,
are for pairs of spins $i$ and $j$ that border a geometric cluster $A$: these spins lie between $A$ and $\overline{A}$.
In our current implementation of the down sweep stage, we compute all pairwise correlations
between the four spins that are on the corners of each rectangular region.
The results of this computation include correlations between all pairs of neighboring spins, as these
are on the corners of the region around a single Kasteleyn city (a $1\times 1$ region).

\subsection{Description of the down sweep stage}\label{subsec:down}
The down sweep stage uses as initial data the cluster
matrices $U_A$ for each node $A$ of the tree found during the up sweep.
As in the FIND method \cite{FIND}, this initial set of cluster matrices is then used to calculate
cluster matrices $U_{\overline{A}}$ for the complementary (i.e., exterior) regions $\overline{A}$.
These matrices encode the boundary correlations of dimer matchings resulting from summing
matchings over the portion of the sample {\em surrounding} a geometric region $A$.
This is to be compared with the cluster matrix $U_{A}$ which contains information about dimer correlations between its
boundary nodes resulting from summing all of the dimers within the region $A$.

\begin{figure}[h]
\centering
\includegraphics[width=3.0in]{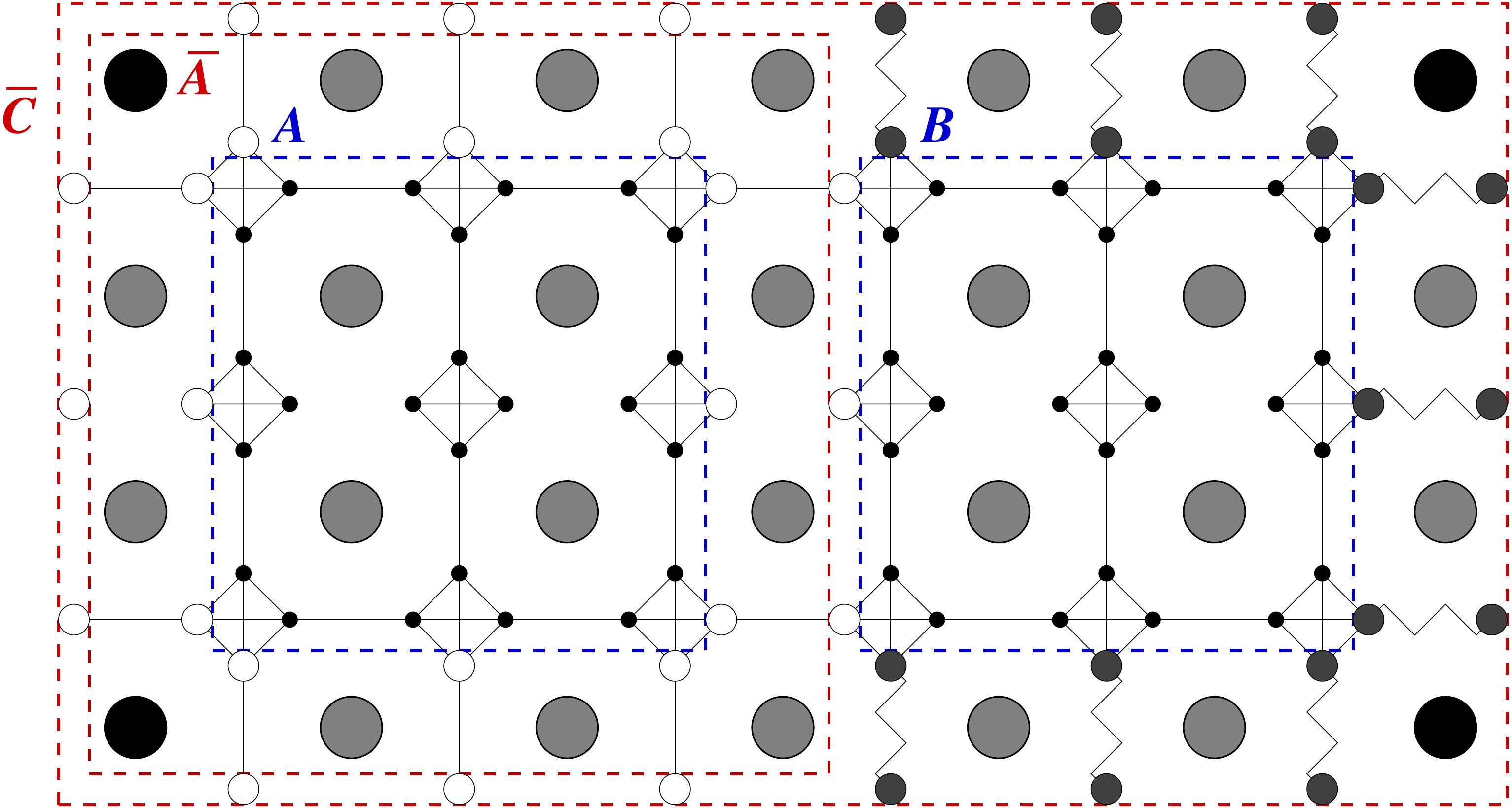}
\caption{(color online) Diagram of a sample merging of cluster matrices in the down sweep stage for the
regions indicated in \figref{fig:sep}. The gray and black larger circles indicate the
locations of the Ising spins in the original square grid. (The black spins are the corner spins for region $C$). The matrix $U_{\overline{C}}$
describes the dimer correlations between the nodes that touch the outer (red) dashed line labeled $\overline{C}$. This matrix
sums over correlations external to the spins shown. This cluster matrix $U_{\overline{C}}$ is merged with
$U_B$ by eliminating the edges shown by the jagged solid lines. The region $B$ is indicated by the dashed
(blue) square on the right of the diagram. The result of the merger is the matrix
$U_{\overline{A}}$ describing correlations among the nodes on the boundary of $\overline{A}$,
which is shown by the labeled square (red) dashed line.
}
\label{fig:downsweep}
\end{figure}

At the highest level, where there are two regions $A^*$ and $B^*$, $U_{\overline{A^*}}=U_{B^*}$ and
$U_{\overline{B^*}}=U_{A^*}$, up to permutations of rows and columns due to differing indexing of the
boundary nodes, as $A^*$ is exterior to $B^*$ and $B^*$ is exterior to $A^*$.
The complementary matrix $U_{\overline{A}}$ for a region $A$ at a lower level is computed by
merging $U_{\overline{C}}$ with $U_B$, where $C$ is the parent region for the siblings $A$ and $B$.
The matrices $U_{\overline{C}}$ and $U_B$ are placed into a larger matrix and the edge weights for those
edges whose ends are shared by these two boundaries are included. Those edges shared by $\overline{C}$ and $B$
are eliminated by Pfaffian elimination
and what remains is the cluster matrix $U_{\overline{A}}$, indexed by the nodes adjacent to $A$.
The entire tree is descended in this fashion, thus generating
complementary matrices and correlations between corner
spins for each region in the geometry tree $T$. A step of this process is diagrammed
in \figref{fig:downsweep}.

As the $U_{\overline{A}}$ are computed, the clusters $U_A$ and $U_{\overline{A}}$ can be merged via
Pfaffian elimination. By comparing the results found using different signs
for the connecting edge weights,  the spin-spin correlations on the original lattice can be computed.
To explain this computation of correlations,
we continue to suppose that domain walls are defined using an all spin up reference configuration $S^r=S^+$, so that
neighboring Ising spins of opposite sign are separated by a domain wall.
Then the Boltzmann weight $e^{-\beta \mathcal{H}(S)}$ of a given spin configuration $S$ is equal to
the product $c^r \Pi_{e\in X(S)} w_e$ of all weights $w_e$ of edges $e$ that make up the domain wall set $M(S,S^r)$ on
the dual graph with $c^r =e^{-\beta\mathcal{H}(S^r)}$, which is a sample and reference state dependent constant .
Consider two spins located at $i$ and $j$ in $G$.
In a given spin configuration $S$, the spins are separated by either an even number or
and odd number of domain walls in $M$.
The spins have equal orientations, $s_i = s_j$,
if and only if a path in $G$ between the two spins crosses an even number of domain walls.
So the correlation function can be found from the average parity of domain walls between the two spins $i$ and $j$.

Given a choice of couplings $J_{ij}$ with chosen boundary conditions and temperature,
let the equilibrium fraction of spin configurations $S$ with $s_i=s_j$ ($s_i\ne s_j$) be
given by $P(s_i = s_j)$ [respectively, $P(s_i\ne s_j)$]. Let $i\rightarrow j$ indicate a path of length $\ell$ between $i$ and $j$ built up of
nearest neighbor pairs $(i,k_1),(k_1,k_2),\ldots,(k_{\ell-1},j)$. For $\ell=1$, the path is just the single bond $(i,j)$.
The partition function under the constraint that $s_i \ne s_j$ is
\begin{eqnarray}
Z_{s_i\ne s_j}=\sum_{S | s_i\ne s_j }e^{-\beta \mathcal{H}(S)}
\end{eqnarray}
and can be represented as the restricted sum over matchings $M$ in $G_D^*$
\begin{multline}
Z_{s_i\ne s_j}\\=\sum_{\{M|i\rightarrow j\ \mathrm{crosses\ odd\ \#\ edges\ in\ }M\}} \epsilon(P) \prod_{e\in M} w(e)\ .
\end{multline}
In this formula, the sum over matchings is understood to be restricted to edge choices that obey the restrictions described below
\myeqref{eq:PfK} and $\epsilon(P_M)$ is the sign of the permutations in the listing of the nodes in the matching $M$, so that the many-to-one mapping
of dimer coverings to spin configurations is effectively turned into a one-to-one mapping by cancellation of oppositely signed terms.
The restriction is to matchings such that the bonds for Ising spin pairs in the path $i\rightarrow j$ are crossed by the dual edges in the matching $M$
an odd number of times. Note that this restriction is independent of the exact path $i\rightarrow j$ and only depends on the endpoints $i$ and $j$.
A similar correspondence (with the sum over an even number of crossings)
holds for expressing $Z_{s_i=s_j}$, with $Z_{s_i=s_j}+Z_{s_i\ne s_j}=Z$.

\begin{figure}[h]
\centering
\includegraphics[width=3.0in]{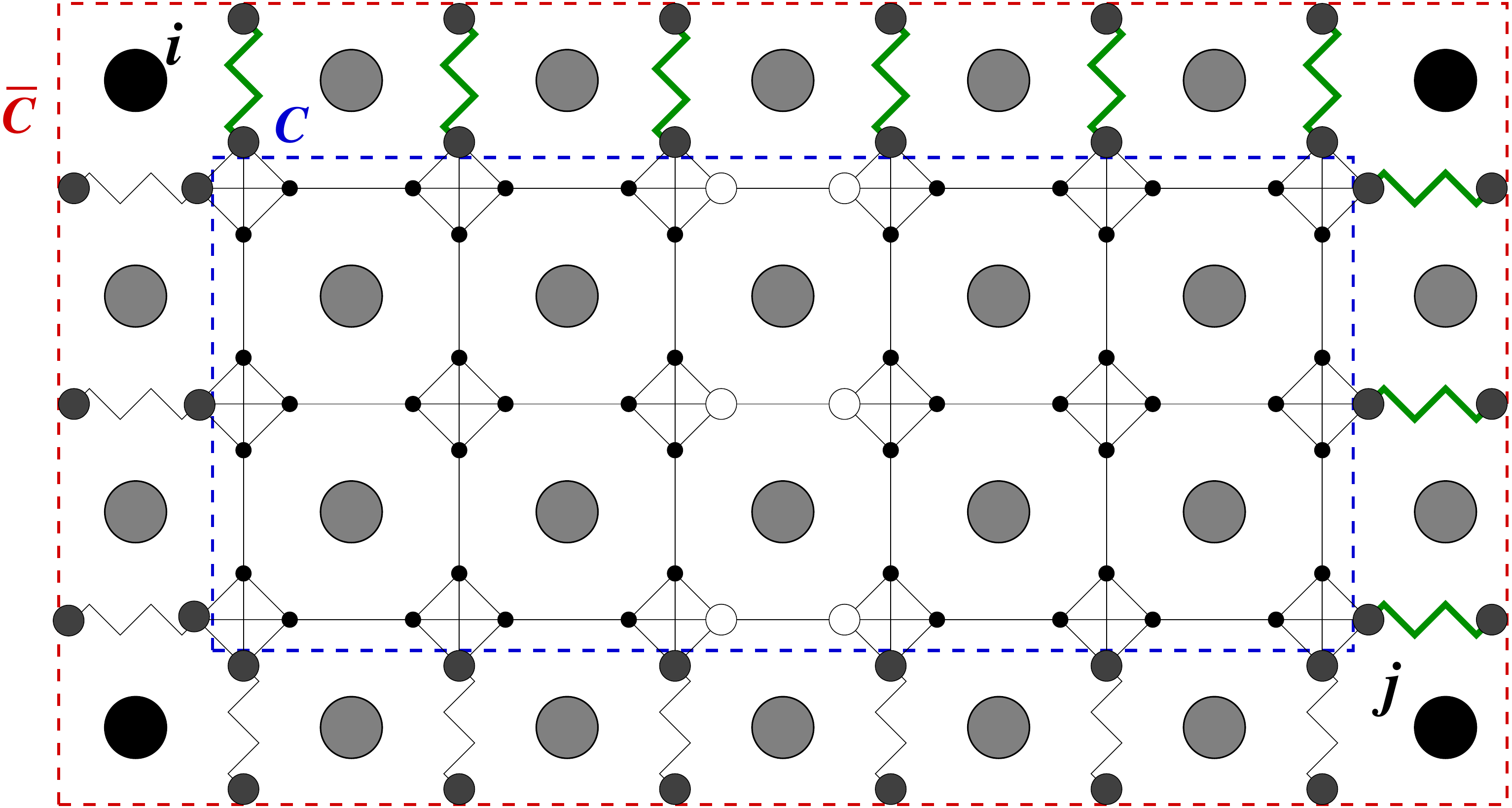}
\caption{(color online) Diagram of a calculation of a correlation between
spins $i$ and $j$. This calculation
uses information computed during the both the up sweep and down sweep stage.
The matrices $U_{\overline{C}}$ and $U_C$ are merged using
the edge weights on the jagged lines. These edges form the set $W$.
Two mergings are calculated: one for all positive weights for the edges
in $W$ and one where the weights are negated for the thicker (green) edges.
This gives two Pfaffians for the whole sample. The first
Pfaffian has positive contributions from configurations with either an odd or even number of dimer
choices (i.e., domain walls) between spins $i$ and $j$. This is the partition function for the whole
sample. The second Pfaffian is the difference between the partition function
constrained to have an even number of domain walls between $i$ and $j$ and the partition
function constrained to have an odd number of domain
walls between $i$ and $j$.
}
\label{fig:corr}
\end{figure}

To compute these restricted partition functions, we compare $\Pf(K)$ with $\Pf(K_{i\rightarrow j})$, where
$K_{i\rightarrow j}$ is a modified Kasteleyn matrix. All weights for the
edges in $G_D^*$ that cross the chosen path $i\rightarrow j$ contained in $G$ are negated in this modified matrix.
That is, the elements of $K_{i\rightarrow j}$ are the same as those in $K$ except
where an edge $e\in G_D^*$ is crossed by the path $i\rightarrow j$: in that case, the weight $w_e$ in $K$
is replaced by $-w_e$ in $K_{i\rightarrow j}$.
These negations reverse the signs of the weights of matchings $M$ with an odd number of edges of that cross
$i\rightarrow j$ while maintaining the sign of matchings with an even number of edges of $M$ crossing that path.
To efficiently carry out the computation of $\Pf(K_{i\rightarrow j})$,
the path $i\rightarrow j$ is chosen to cross edges that connect a cluster $A$ to its complement $\overline{A}$.
That is the path connect spins that lie between $A$ and $\overline{A}$.
An example showing the negated dual edges is shown in \figref{fig:corr}.

These correspondences allow us to write the spin correlation function for a planar Ising model in the form
\begin{eqnarray}
\langle s_i s_j \rangle &=& P(s_i=s_j)-P(s_i\ne s_j) \nonumber\\
& = & Z^{-1}\left[Z_{s_i = s_j} - Z_{s_i \ne s_j}\right]\nonumber\\
&=  &\frac{1}{\Pf(K)}
     \left[\sum_{\{M|\mathrm{even}\ i\rightarrow j\}}
         \epsilon(P_M)\prod_{e\in M}w(e)- \right. \nonumber \nonumber\\
         & &
      \left. \sum_{\{M|\mathrm{odd}\ i\rightarrow j\}}\epsilon(P_M)\prod_{e\in M}w(e)\right] \nonumber\\
      & = & \frac{1}{\Pf(K)}\Pf(K_{i\rightarrow j}) \nonumber\\
       & = & \frac{\Pf[M_0(W_{i\rightarrow j}, U_A, U_{\overline{A}})]}{\Pf[M_0(W,U_A,U_{\overline{A}})]},
\label{eq:corr}
\end{eqnarray}
The modified list of weights $W_{i\rightarrow j}$ is the set of weights with negated values for all dual edges crossed by the path
(any path) from $i$ to $j$, i.e., $i\rightarrow j$, where again $i\rightarrow j$ lies between $A$ and $\overline{A}$.
Note that $\Pf(K) = \Pf[M_0(W)]z(A)z(\overline{A})$ and $\Pf(K_{i\rightarrow j})=\Pf[M_0(W_{i\rightarrow j)}]z(A)z(\overline{A})$;
the cancellation of the common factor $z(A)z(\overline{A})$ gives the last step in the above equation.

This representation of the spin correlations in \myeqref{eq:corr} defines the procedure for their computation.
Correlations between two spins that lie between a region $A$ and and its complement $U_{\overline{A}}$ are
computed
by merging the two matrices $U_A$ and $U_{\overline{A}}$ once using the original weights and again using
the modified (partially negated) weights.
The ratio of the two resulting Pfaffians gives the spin-spin correlation value.
We note that a different approach has been used to compute correlation functions at
$T=0$ \cite{PoulterBlackman}, where paths between frustrated plaquettes are the basis of the
representation in the ground state.
For the example shown in \figref{fig:corr},
the correlation function is calculated between the two spins diagonally opposite (top left
and bottom right) between the outer and inner set of nodes, whose correlations are given
by $U_A$ and $U_{\overline{A}}$.
The choice of signs for $W$ could also be modified to compute multispin correlations.

The calculation of correlations is simplest for planar graphs (Ising models with open or fixed
boundary conditions).  If periodic boundary
conditions are to be used, four different mergers of the two top level matrices $U_{A^*}$ and $U_{B^*}$
are computed.
These mergers are
computed for all possible pairings of negative or positive weights for bonds that connect the top row to the bottom
row of cities or the rightmost column to the leftmost column of cities, as justified in 
Sec.\ \ref{subsec:up} and Ref. \cite{ThomasMiddletonSampling}.
The descent of the tree for each is carried out starting from each of these four choices.
There will then be four complementary cluster matrices, $U^{\pm,\pm}_{\overline{A}}$, for each geometrical region $A$.
Each complementary cluster matrix will have its own four partial Pfaffian factors $z^{r,s}(\overline{A})$.
A spin-spin correlation for periodic boundary conditions (as given by $K^{++}$) is then the ratio of two weighted sums.
The weighted sum in the denominator is the total partition function divided by $z(A)$.
The sum in the numerator is the same linear combination of the Pfaffians but with the weights $W$ negated on edges
that cross the path $i\rightarrow j$ (i.e., using the weights $W_{i\rightarrow j}$). The partition function for periodic boundary
conditions is $Z^{P,P}=\frac{1}{2}\sum_{(r,s)} \Pf(K^{r,s})$. The
factor $z(A)$ is common to all terms $\Pf(K)=z(A)z(\overline{A})\Pf[M_0]$
in the weighted sums and so can be cancelled out.
This gives the result that the spin-spin correlation function on a periodic lattice is the ratio
\begin{eqnarray}\label{weightedCorrs}
\langle s_i s_j \rangle = \frac {\sum_{(r,s)\in\{(\pm,\pm)\}} \Pf[M_0(W_{i\rightarrow j},U_A,U^{r,s}_{\overline{A}})]z^{(r,s)}(\overline{A})}
								{\sum_{(r,s)\in\{(\pm,\pm)\}} \Pf[M_0(W,U_A,U^{r,s}_{\overline{A}})z^{r,s}(\overline{A})]}\ .
\end{eqnarray}
So the correlation function computations,
which require the evaluation of two Pfaffians on a planar graph, require $8$
Pfaffians on a torus for each pair of spins. The computation time for the correlation functions for the corner spins on
all regions in practice requires about 10 times the amount of computing time as finding only the partition function $Z$.

\section{Sampling}\label{sec:sampling}

Exact sampling methods select independent configurations according to their probability in the whole sample space.
We consider here the problem of generating a sample configuration of a system with probability proportional
to the Boltzmann weight $e^{-\beta \mathcal{H}}$.
As a contrast with direct sampling, consider Markov chain Monte Carlo (MCMC) methods.
In an MCMC method, 
a sequence of configurations is generated by randomly chosen updates; if the update choices obey
detailed balance and can reach all possible configurations, in the limit of large times
this sequence will generate sample configurations from the Boltzmann
distribution \cite{NewmanBarkema}. The number of Monte Carlo updates needed for the approach to fair sampling
is often unknown and can be very long.
However, MCMC methods can generate exact sampling if coupling from the past \cite{cpftp} can be used to
guarantee fair samples (but not necessarily fast mixing times). However,
no known coupling methods are practical for Ising spin glass models at low
temperatures \cite{KrauthCouplingLowT}. Markov chain Monte Carlo methods are of course of great practical
use, but the availability of exact sampling in some cases provides for a very useful comparison and
the potential for much more rapid calculations for large glassy systems.

The direct sampling methods we use \cite{ThomasMiddletonSampling} to generate random Ising spin configurations
are based on the mapping between dimer and Ising model configurations and on dimer sampling methods
\cite{Wilson} that use nested dissection. By directly selecting a random matching on the decorated dual graph $G_D^*$ with the
proper probability, we fairly select a set of relative domain
walls and hence the relative orientations of the spins on the original lattice.
We report here on a modification of the method used in Ref.~\cite{ThomasMiddletonSampling}; here we
use the edge separators $W$ \cite{FIND} described in Sec.~\ref{sec:main}
rather than a node separator \cite{Wilson}.
This modification significantly speeds up the sampling algorithm for the Ising model,
as the dimension of the matrices to be factorized are reduced
by a factor of three from those used in Ref.~\onlinecite{ThomasMiddletonSampling}.
The implementation of the algorithm is also simplified.

\begin{figure*}[h]
\centering
(a)\ \includegraphics[width=1.4in]{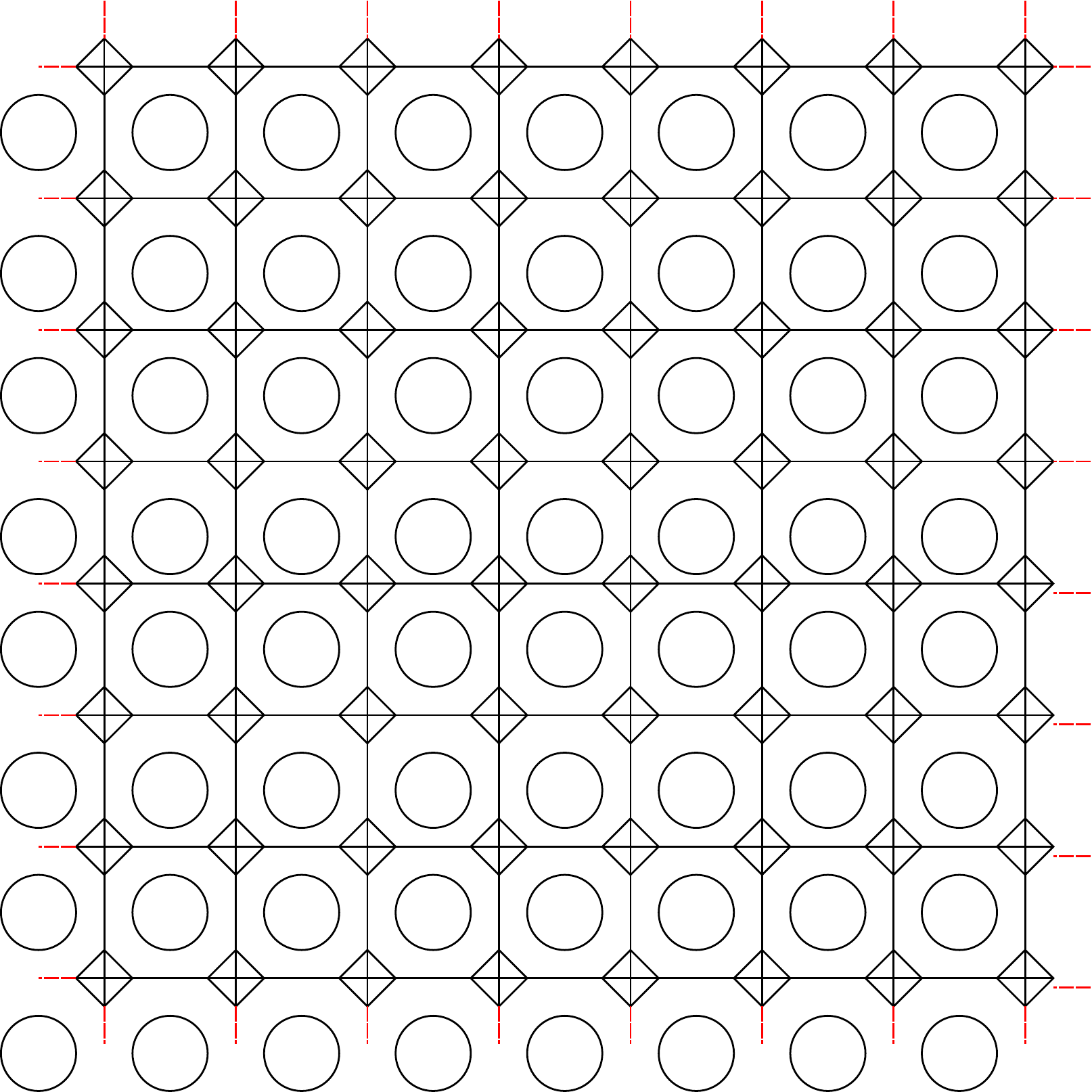}
\ (b)\ \includegraphics[width=1.4in]{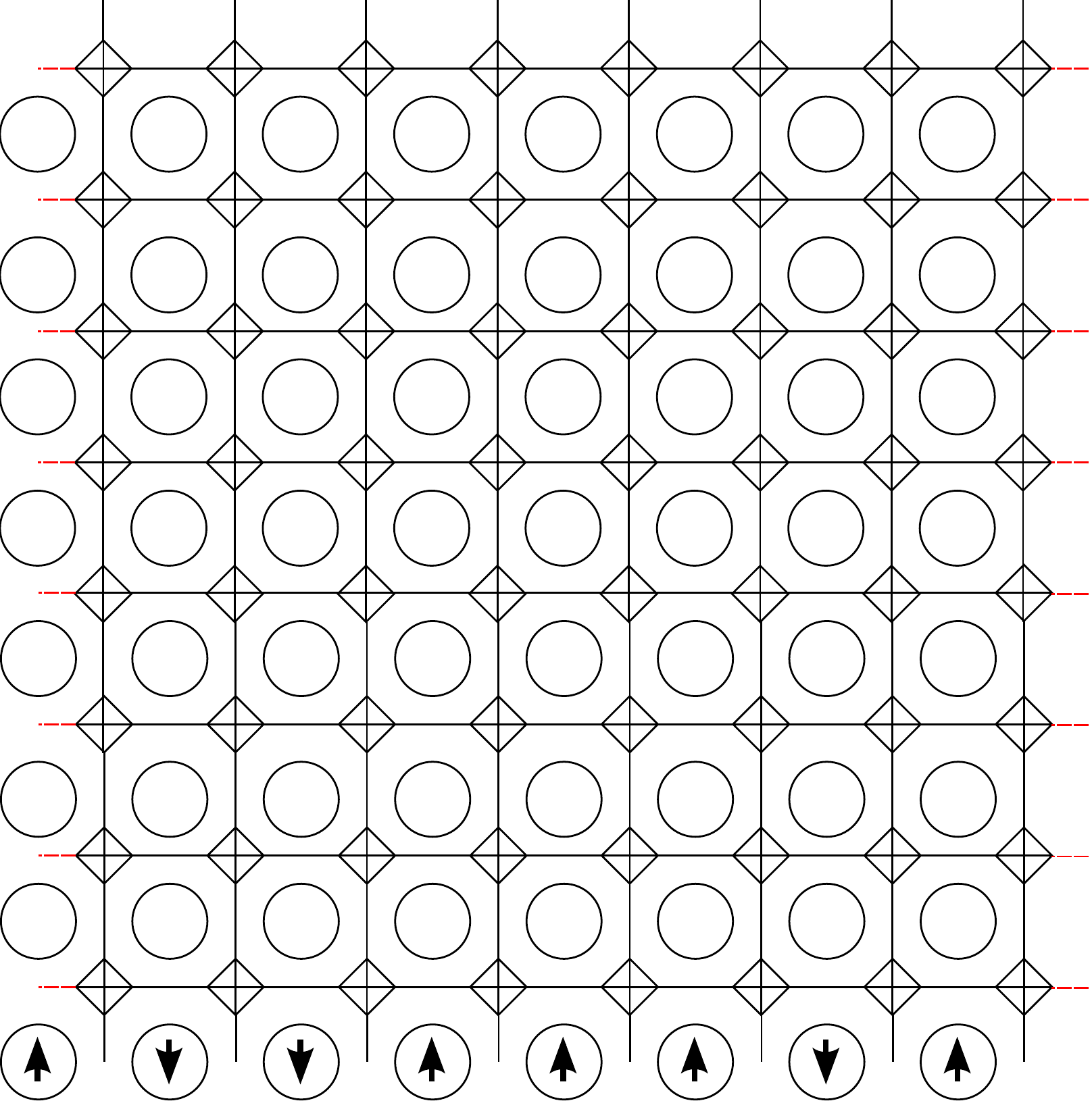}
\ (c)\ \includegraphics[width=1.4in]{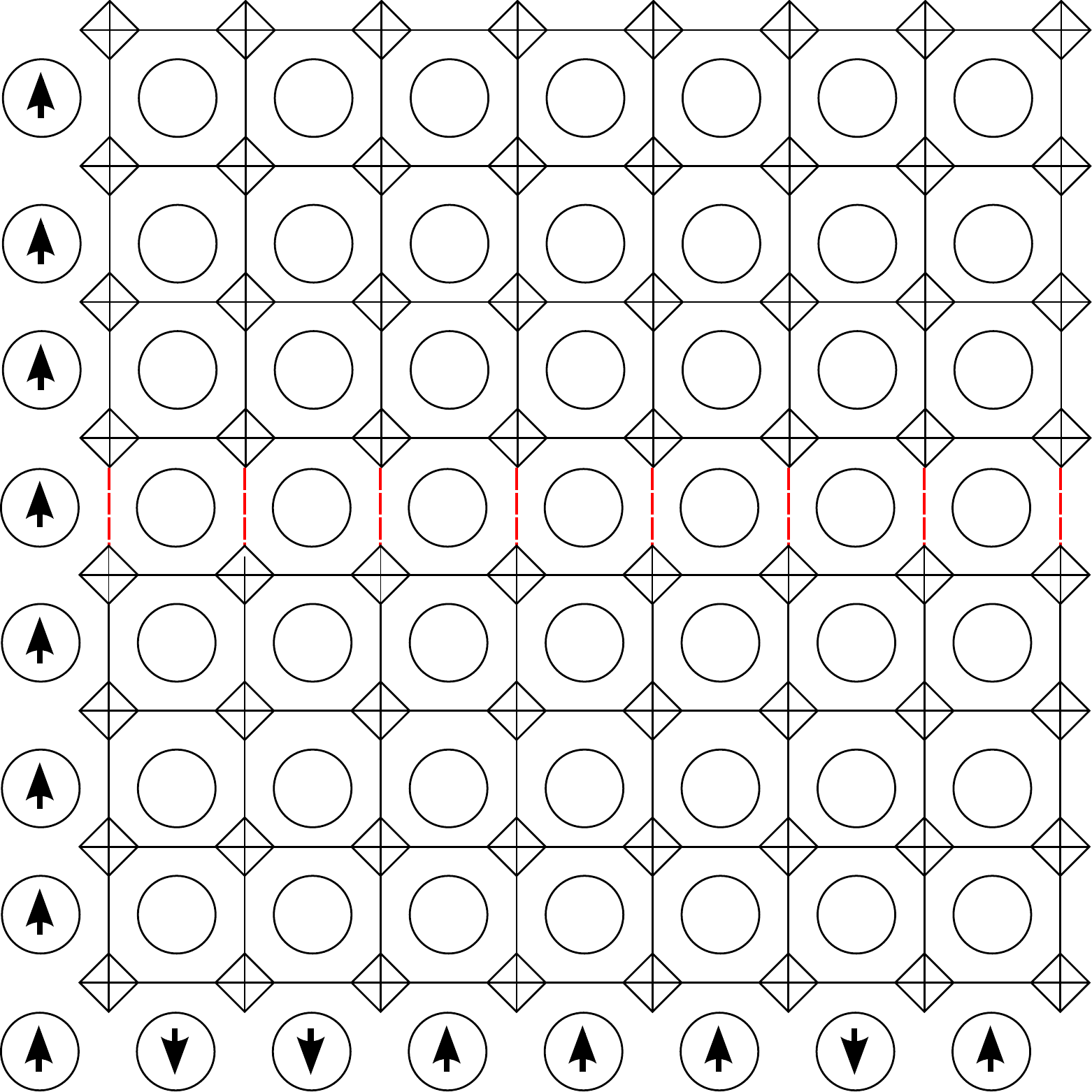}
\ (d)\ \includegraphics[width=1.4in]{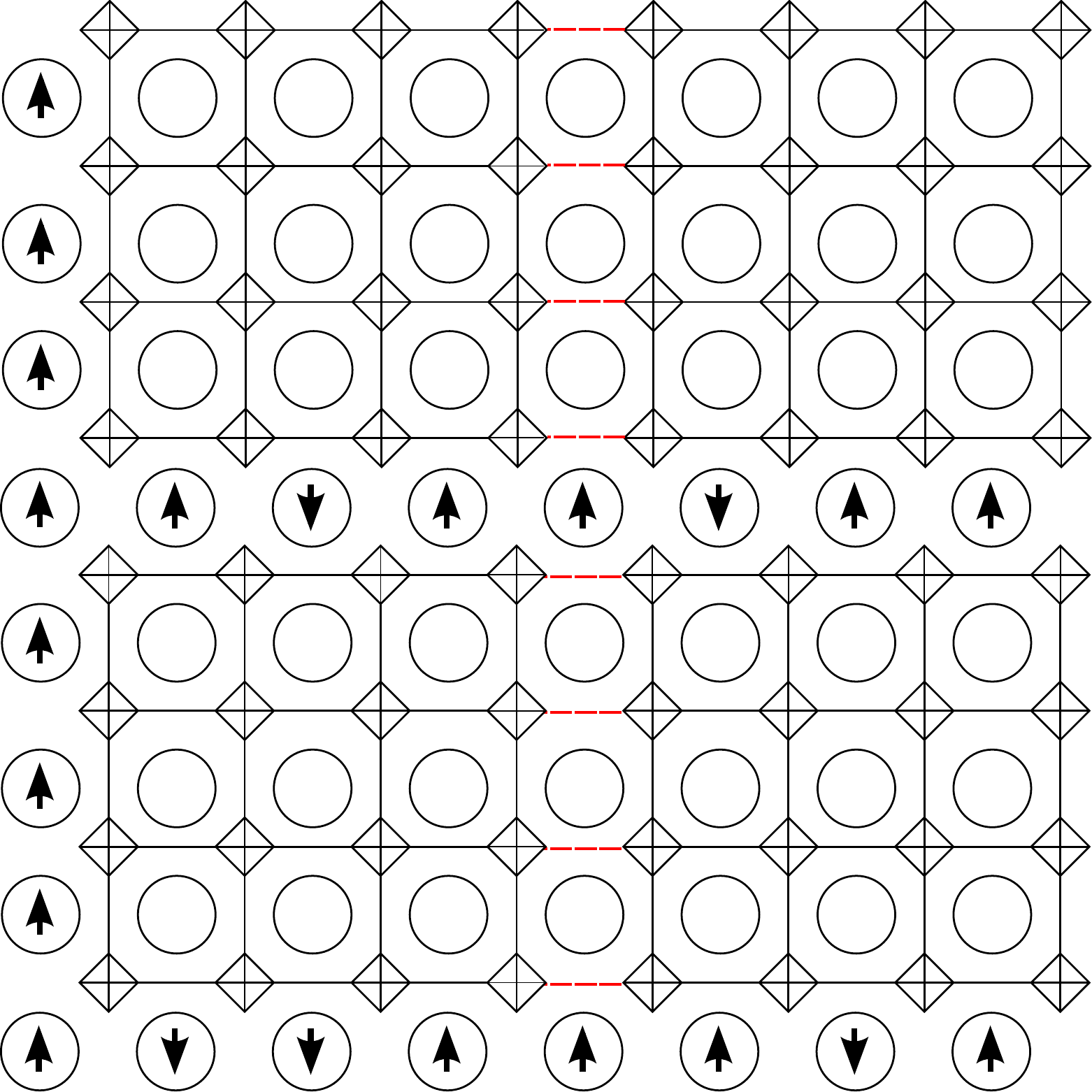}\\\ \\
\ (e)\ \includegraphics[width=1.4in]{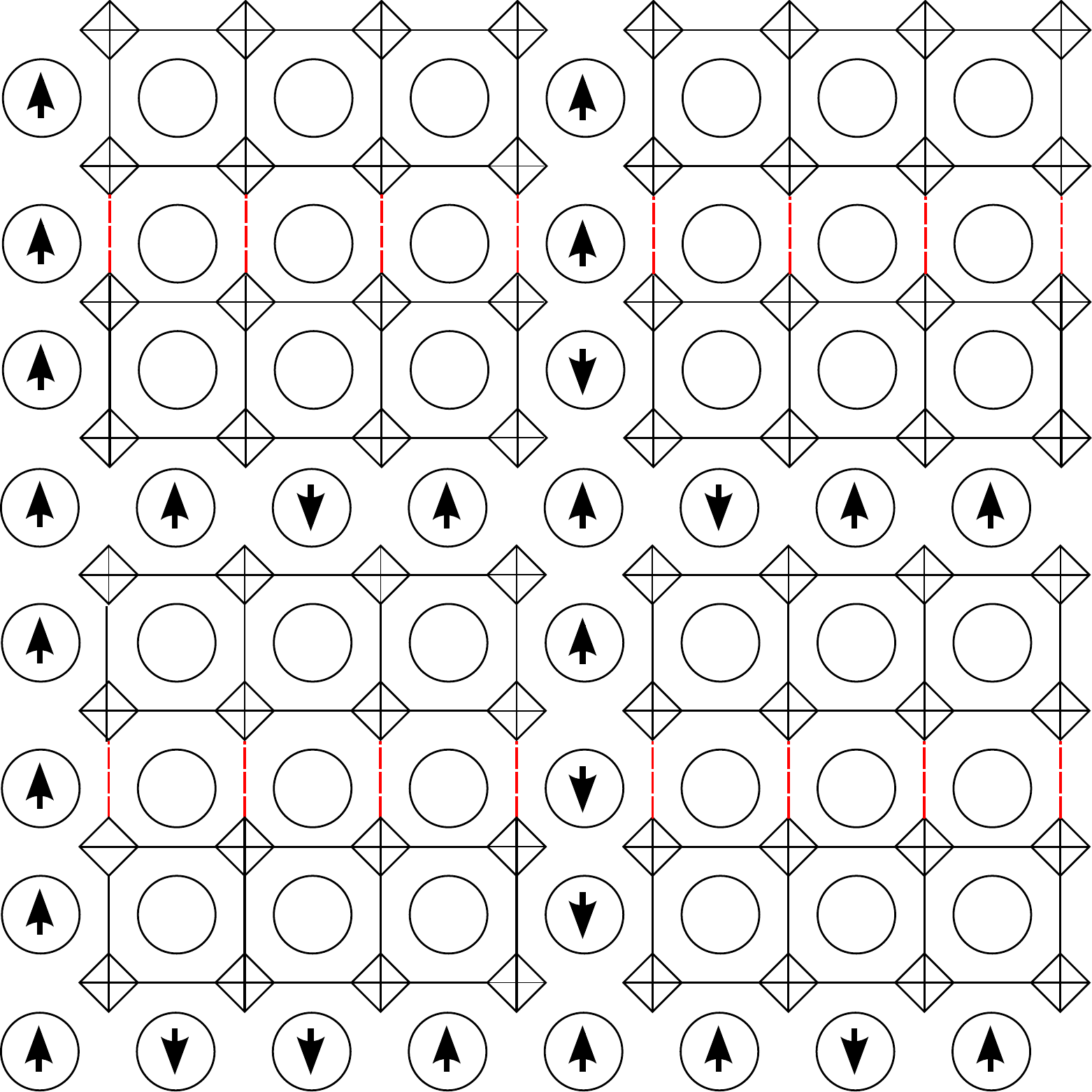}
\ (f)\ \includegraphics[width=1.4in]{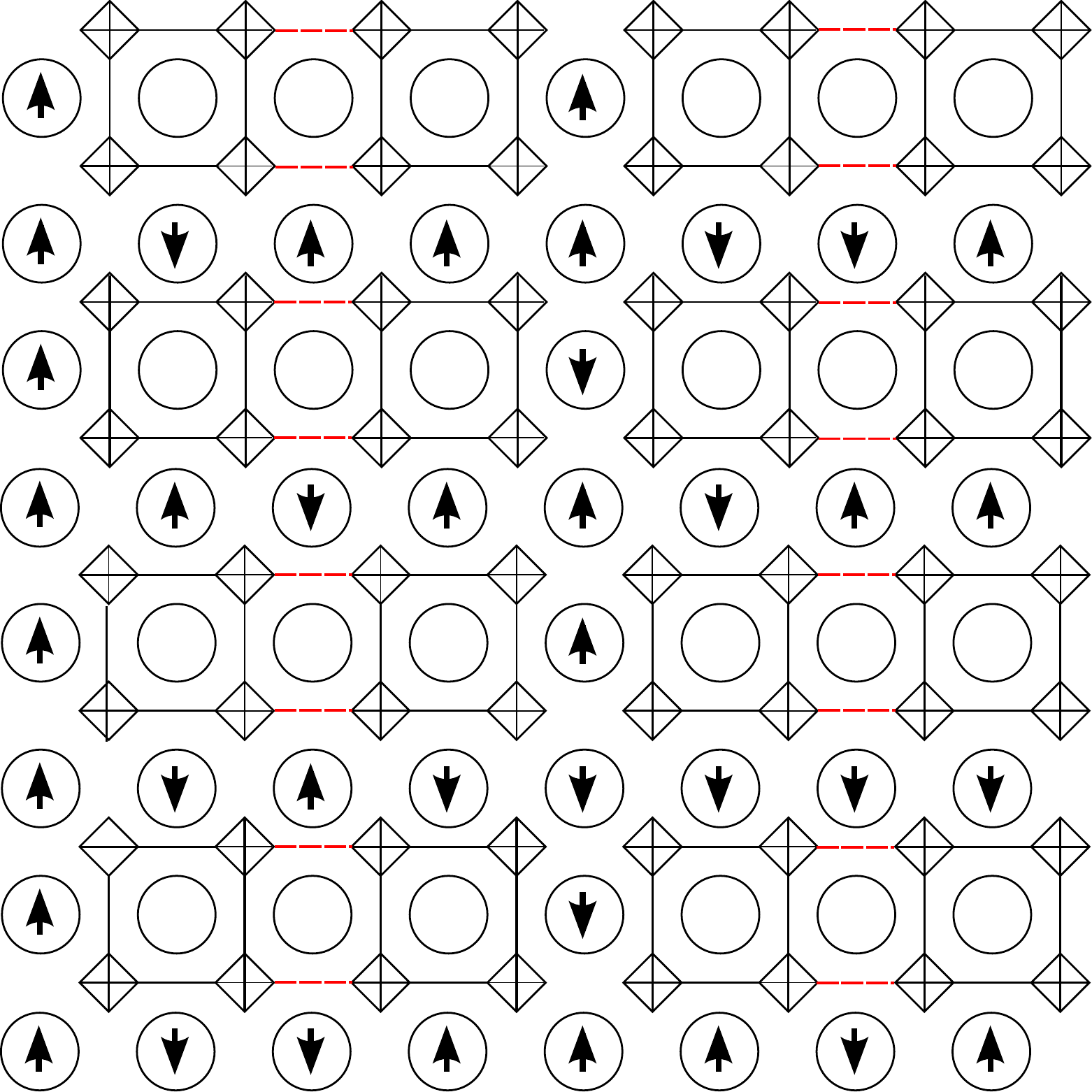}
\ (g)\ \includegraphics[width=1.4in]{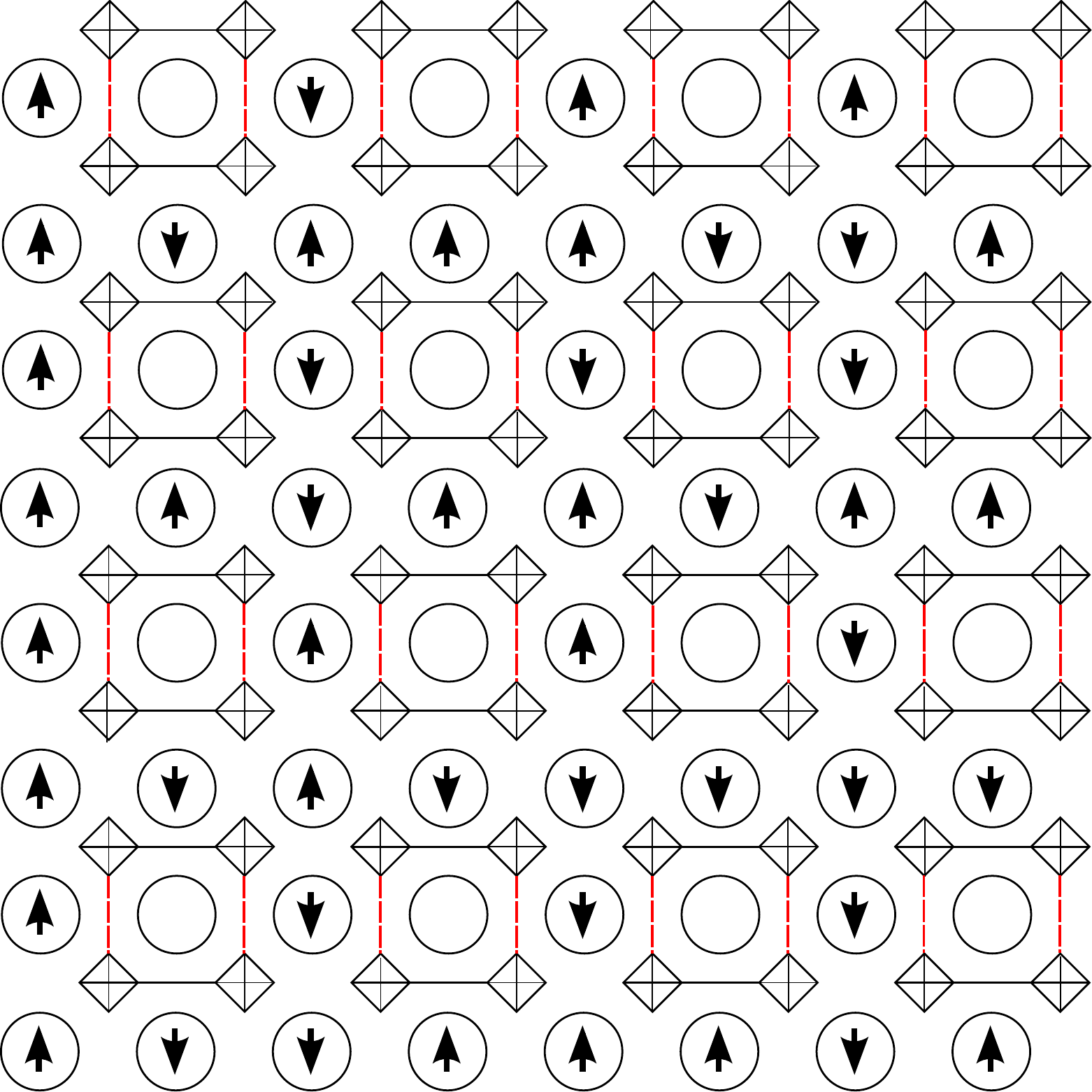}
\ (h)\ \includegraphics[width=1.4in]{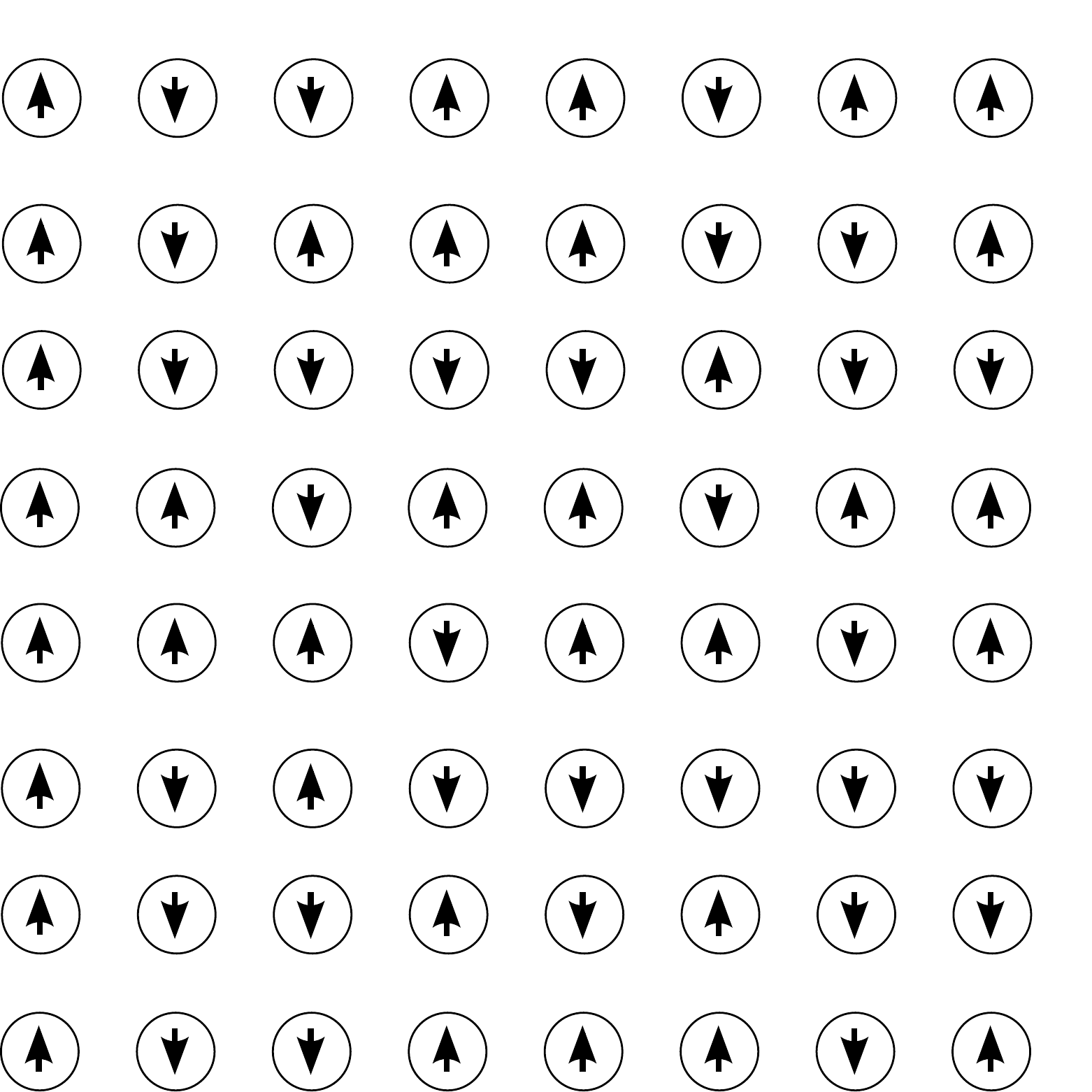}
\caption{(color online) Steps for exact sampling on a periodic
Ising spin lattice of size $L_x\times L_y = 8\times 8$. Empty circles indicate unknown
spins. Arrows indicate assigned spins: $\uparrow$ for $s=+1$ and $\downarrow$ for $s_i=-1$.
Separators $W$ are drawn as dashed light (red) lines.
The regions of the graph $G_D^*$ used at each stage are shown
by solid darker lines. (a) The first region $C^{*}$ is shown. The separator $W$
(these wrapping edges are drawn as separated half edges) connects $C^*$ to itself. Given a seed spin, deciding
which edges in $W$ are in the matching $M$ along the bottom row fixes the spins for step (b).
(b) The spins decided in (a) fill the lowest row. The remaining separator along the column is to be filled in
for the start of the next step. (c) In all subsequent steps, including this step, the boundary conditions are fixed.The region $C^*$
is separated into the lower half $A^*$ and the upper region $B^*$. Choices are made for
the dual edges connecting $A^*$ to $B^*$. (d) Three edges are chosen for each of the two
separators to fix the 3 spins for each region pair. (e) Four separators are used to set 12 spins. (f) Eight separators
are used to set eight spins. (g) In this final stage, there are 16 separators. One edge choice is made for each,
fixing the remaining undecided spins.
(h) The final spin assignment.
}\label{fig:samp}
\end{figure*}

As in the computation of the partition function and correlation functions,
the direct sampling calculations rely on a geometric dissection.
The tree used for sampling differs some from that described in Sec.~\ref{sec:main} for
computing partition functions and correlation functions.
For sampling configurations with periodic boundary conditions, we start this modified dissection with a cluster
$C^*$ which is formed by joining the two system halves $A^*$ and $B^*$ along a single line of spins.
The cluster $C^*$ includes all of the nodes in the decorated dual graph $G_D^*$, but does not include the edges
at the top or right that connect the top row of nodes to the bottom row or the right column to the left column.
These edges that are left out are those used to complete the periodic boundary conditions. See \figref{fig:samp}(a) for
a drawing of $C^*$ and the initial separator.
All of the edges internal to the region $C^*$ are
contracted out by Pfaffian elimination in an up sweep to give the cluster matrix $U_{C^*}$.
This matrix $M(C^*)$ is used in the first stage of spin assignment.
In this first stage, the
Ising spins that form the bottom row of the sample are chosen.
As the probability distribution is symmetric with
respect to global spin reversals, we can simply fix an initial spin, the spin at the lower left, to have the value $+1$.
The orientation of the remaining spins that lie along the bottom row of $C^*$ are then
assigned sequentially first along the bottom row. The spins along the left column are then
assigned. This assignment is based
on the probabilities of domain walls separating neighboring spins in the bordering row and column.
These probabilities are found by effectively computing the correlation functions between spins in the lower row and left column.
Note that, in principle, any order of spin assignment for these outer border spins could be used.
It is possible that numerical stability might be improved by choosing an alternate order of spin assignments; we chose the 
nearest neighbor sequence for simplicity.

Once all spins around the boundary of the sample are fixed, the process becomes simpler.
Spin assignments are decided at finer scales by descending the tree recursively.
In each subsequent step, the spins surrounding a region $C$ have been fixed by prior assignment.
The probabilities of domain wall sections crossing between the spins lying between two child regions $A$ and $B$
are computed.
The spins between the two child regions $A$ and $B$ are then assigned by using these probabilities of relative domain
walls. As the assignments are made, the probabilities for remaining parts of the separator are updated.
These newly assigned spins then form the boundaries for the child regions of $A$ and of $B$.
These steps are shown for a sample spin assignment in \figref{fig:samp}.

The iterative assignment of spins along the separators uses the inverse of a cluster
matrix to compute correlation functions. The spins are randomly chosen according to these correlation functions.
An essential part of this approach is that when a spin is fixed by such a choice, the inverse of the cluster matrix
can be updated efficiently and incrementally \cite{Wilson,MartinRandall}.
This incremental update makes the sampling procedure running time for selecting a single
spin configuration proportional to the time of computing the partition function
(though with a larger prefactor). A summary outline of the procedure is presented in Sec.~\ref{SamplingOutline}; the next sections
Sec.~\ref{DWprob} and Sec.~\ref{SM} give more details of the algorithm.

\subsection{Computing domain wall probabilities}\label{DWprob}

Given two regions $A$ and $B$ and their edge separator $W$, the domain wall probabilities
are calculated using the inverse of $K_W$, the Kasteleyn matrix that incorporates the effects of
both the values of the boundary spins surrounding $A\cup B$ and the weights on the edges in $W$.
Note that boundary spins around the region $A\cup B$ are taken to be fixed, except at the highest level.
These fixed spins affect the weights of the intercity edges at the boundaries of $A$ and $B$.
These weights are computed for a reference configuration where the spins at the boundary of $A$ and $B$ are fixed
to the values decided at the higher levels of the tree. We can continue to use all spins set to $s_i=+1$ for the spins interior
to $A$ and $B$. It follows that if a spin $j$ neighbor to $i$ in the region $A$ is fixed to be $s_j=-1$, the weight $w(e)$ for the
edge $e$ crossing the bond $\langle i j \rangle$ is set to be $\exp(2\beta J_{ij})$.  If the boundary spin is fixed to $s_j=+1$,
then the usual weight is used, $w(e)=\exp(-2\beta J_{ij})$.
At each stage of the spin assignment, then, we recompute the matrices $U_A$ and $U_B$ using these weights that
depend on the boundary spins for $A\cup B$.
The matrix $K_W$ is
found by collecting $U_A$ and $U_B$ into a single matrix and then linking the matrices using the edge
weights that connect $A$ and $B$.
Matrix inversion using Pfaffian factorization is then used to compute the matrix $K_W^{-1}$.

The inverse matrix $K_W^{-1}$ allows for the simple calculation the probability of any given separating edge
being part of a domain wall.
These calculations Eqns.~(\ref{eq:P},\ref{eq:Pper})
use the Pfaffian analog of the Jacobi determinant identity, which states that
\begin{eqnarray}\label{eq:Jacobi}
\frac{\Pf(U_{k,l})}{\Pf(U)}=\pm\Pf(\{U^{-1}\}_{k,l}),
\end{eqnarray}
where  $U_{k,l}$ is the matrix with rows and columns $k$ and $l$ removed. The notation $\{U^{-1}\}_{k,l}$ indicates
the $2\times 2 $ submatrix of $U^{-1}$ which is built out of the intersections of rows and columns $k$ and $l$, i.e.,
\begin{eqnarray}
\{U^{-1}\}_{k,l}=\left(\begin{array}{cc} 0 & U^{-1}_{k,l}\\ -U^{-1}_{k,l} & 0 \end{array}\right)\,,
\end{eqnarray}
where $k<l$, so that $\Pf([U^{-1}]_{k,l})=U^{-1}_{k,l}$.
Since the ratio of Pfaffians of Kasteleyn matrices $\Pf(K)$ is the ratio of partition functions $Z$, probabilities
can be computed using the identity \myeqref{eq:Jacobi}.
Let the edge on the decorated dual graph that separate
the neighboring spins $i$ and $j$ have nodes $a$ and $b$. The probability of including an edge $e_{ab}$ as part
of a domain wall that separates
the two spins is then given by the expression
\begin{eqnarray}\label{eq:P}
P(e_{ab}\in M)=|[K_W]_{ab}[K_W^{-1}]_{ab}|\ .
\end{eqnarray}
The result Eq.~(\ref{eq:P}) then follows from the probability
being the product of the weight of the chosen edge and the weight of dimer configurations that don't include nodes $k$ and $l$
(i.e., $\Pf(K_W)_{k,l}$) divided by the total weight $\Pf(K_W)$.
When $W$ is the separator for regions $A$ and $B$ with fixed boundary spins, the $ab$ element of $K_W$
is given by $[K_W]_{ab}=w_{ab}$.
The situation is different for the top level matrix $C^*$, where $W$ ``separates'' $C^*=A^*\cup B^*$ from itself, i.e.,
the edges connect boundary nodes on $C^*$ to each other. In this case, $[K_W]_{ab}=w_{ab}+[U_{C^*}]_{ab}$.
In addition, the probability for selecting an edge that connects $C^*$ to itself is given by a weighted sum over the
four possible boundary dimer orientations,
\begin{eqnarray}\label{eq:Pper}
  P(e_{ab}\in M) =
  \left| \frac
   {\sum_{(r,s)}w^{r,s}_{ab}\Pf(K_{W}^{(r,s)})[K^{(r,s)}]^{-1}_{ab}}
   {\sum_{(r,s)}\Pf(K_{W}^{r,s})}
   \right|
\end{eqnarray}
The equation for domain wall probabilities at the highest level
in the periodic lattice follows from \myeqref{eq:L} and \myeqref{eq:Jacobi} and cancellations of common factors similar to those
that led to the result \myeqref{eq:corr}.

Once the probability of choosing an edge is computed, a random number $y$ is chosen in the interval $[0,1)$ to
decide whether to accept the addition of edge $e_{ij}$ to $M$. If $y<P(e_{ij}\in M)$ the edge $e$
is included in the sampled matching, otherwise
it is excluded.
Given the resulting choice, the matrix $K_W^{-1}$ is then updated by the method described in Sec.~\ref{SM}.
We note here the contrast with methods for dimer covering sampling that are based on
node separators \cite{Wilson,ThomasMiddletonSampling}.
In these methods,
a chosen node that is between regions $A$ and $B$ was matched. The probabilities computed were the
probability of choosing each edge that matched the chosen node, with the sum of these probabilities being unity.
From the Jacobi
identity, the conditional probabilities for these forced node matching could be found without recomputing
all of the elements of $K^{-1}$; these probabilities are given by the Pfaffian of a submatrix that grows
with the number of fixed nodes \cite{Wilson}.
Here, instead, nodes on the boundary may or may not be matched, depending on whether an edge is chosen or not,
so the same approach cannot be used.
Instead, inspired by the approach of Ref.~\cite{MartinRandall}, we update the inverse matrix using the Sherman-Morrison formula.
Note that only $|W|-1$ edges are need be chosen for each separator $W$, as the last choice of an edge is forced by consistency
in the spin assignments (or, equivalently, parity in the dimer covering.)

\subsection{Updating $K^{-1}$ using the Sherman-Morrison formula}\label{SM}

After the
assignment of one spin value, the choice of whether the corresponding edge
is included in the matching is fixed for the remainder of the
calculation; all subsequent bond probabilities along the separator $W$ must be computed conditioned
upon this choice.  This is accomplished by modifying the inverse Kasteleyn
matrix $K_W^{-1}$ for the edge separator.
The Sherman-Morrison formula \cite{MartinRandall} allows for quickly recomputing
the inverse of a matrix when modifications of the original matrix are confined
to one (or a small number) of rows and columns.
Here, we apply this formula to set specific edges of $K_W$ to zero.
One formulation of the Sherman-Morrison formula is
that for any matrix $A$, and row vectors $u$ and $v$,
\begin{eqnarray}
\left( A + u v^T \right)^{-1} & = & \left(A^{-1} -
\frac{A^{-1}uv^TA^{-1}}{1+v^TA^{-1}u}\right).
\end{eqnarray}
In the sampling algorithm, the choice of whether to force the inclusion of an edge or exclude it 
from $M$ modifies the skew-symmetric matrix $K_W$ in two rows and columns at the same time.
When including an edge $e_{ab}$, all elements of $K_W$ in rows and columns $a$ and $b$ are
set to zero, except the elements $[K_W]_{a,b}$ and $[K_W]_{b,a}$. The matching found using $K_W$
must then link $a$ to $b$. In contrast, when the edge is excluded, these two elements $[K_W]_{a,b}$ and $[K_W]_{b,a}$
are set to zero, while the others
in rows and columns $a$ and $b$ are kept unchanged.
If a general matrix $A$ (and hence its inverse $A^{-1}$) is antisymmetric, numerical
stability is enhanced by carrying out both row operations and column operations at the same
time, keeping the resulting matrix antisymmetric as well.
Using skew-symmetry and applying the Sherman-Morrison formula twice gives the inverse of a
matrix modified in two rows and columns as
\begin{eqnarray}
\label{eq:SM_AS}
\left( A + u v^T -(uv^T)^T \right)^{-1} & = &
A^{-1} - \left(
\frac{A^{-1}uv^TA^{-1}}{1+v^TA^{-1}u} \right) 
\nonumber \\ 
& &+ \left(
\frac{A^{-1}uv^TA^{-1}}{1+v^TA^{-1}u} \right)^T.
\end{eqnarray}
If an edge $e_{ab}$ is to be removed by setting $K_{ab}$ to zero,
then one can set $u_a = 1$ and $v_b= -K_{ij}$,
with all other elements of $u$ and $v$ being zero.
Similarly, if edge $e_{ij}$ is to be kept, one can also use $u_k=\delta_{a,k}$ for Kronecker delta $\delta$ 
but set the vector $v$ by $v_k = -K_{bk}$, for $\forall k \neq j$.

Given that that matrix $A^{-1}$ is computed directly only once, at the start of spin assignment along a separator,
a single update procedure per Eq.~\ref{eq:SM_AS} may be carried out in $\mathcal{O}(L^2)$
operations for a separator of $L$ spins.  This is faster than the
$\mathcal{O}(L^3)$ for matrix multiplication of two $L\times L$ matrices
because $A^{-1} u$ and $v^T A^{-1}$ are themselves vectors.
So performing $L$ updates to the matrix can be achieved in $\mathcal{O}(L^3)$
operations.  As the nested dissection produces a separator with $L\propto
\sqrt{N}$ elements, sampling across the separator takes $\mathcal{O}(N^{3/2})$
steps. Summing this cost over all of the needed scales for the separators gives a time
to sample all of the spins scaling also as $\mathcal{O}(N^{3/2})$.

\subsection{Sampling algorithm outline}\label{SamplingOutline}

Given the connection between $K_W^{-1}$ and spin correlations and the Sherman-Morrison method for updating
$K_W^{-1}$ as spins are chosen, the sampling algorithm for periodic
systems can now be directly described in outline form:

\begin{enumerate}
\item Perform the same up sweep steps needed to compute
the cluster matrices $U_{A^*}$ and $U_{B^*}$. These are the same steps needed to compute the partition function $Z$
but without the final merger.
\item Merge $U_{A^*}$ and $U_{B^*}$ along the line of spins through the middle of the
sample that separates the regions $A^*$ and $B^*$. This is done
by placing $U_{A^*}$ and $U_{B^*}$ into a larger matrix and filling in the values of weights $w_{ab}$ along this
line. This gives the cluster matrix $U_{C^*}$ which is indexed by nodes along the bottom, top, left, and
right rows of the sample.
\item For each of the four global dimer orientations $\pm\pm$, fill in the weights that complete the 
torus, using signs for the weights given for each choice $(r,s)\in{\pm\pm}$. This gives four matrices $U_{C^*,\pm\pm}$.
Compute the inverse matrices $U^{-1}_{C^*,\pm\pm}$ using Pfaffian elimination (\myeqref{eq:PfInverse}).
\item For each edge $e$ that connects the top of $C^*$ to the bottom of $C^*$:
  \begin{enumerate}
  \item Compute the probability that the edge $e$ is occupied using
  \myeqref{eq:Pper}.
  \item Apply \myeqref{eq:SM_AS} to update $U^{-1}_{C^*,\pm}$, using the vectors $u$ and $v$ that modify
   $U_{C^*}$ so as to force the chosen occupation value of the current edge $e$.
  \end{enumerate}
\item Compute two new cluster matrices, $U_{C_1^*}$ and $U_{C_2^*}$ which have as boundaries the left and right
columns of $C^*$, using the fixed values of the spins in the bottom row chosen in the previous step. Then use a modified
form of \myeqref{eq:Pper} that sums only over $s$, not both $r$ and $s$, to compute probabilities for edges along the
column at the left/right boundary of the sample. After selecting each edge, use \myeqref{eq:SM_AS} to update
$U^{-1}_{C_1^*}$ and $U^{-1}_{C_2^*}$.
\item Use the edge choices, which give portions of relative domain walls, to assign Ising spins around the border of $C^*$.
More specifically, if an edge is chosen to belong to the matching $M$,
the sign of the spin differs on either side of the edge, while if a spin was not chosen, the sign
of the two spins on either side is the same.
\item Sampling is now carried out recursively for the subregions,
given these fixed boundary conditions around the border of the sample.
This is carried out first for the spins lying on the central dividing line
between $A^*$ and $B^*$, and then for the spins lying between their child regions, etc.,
until all spins are assigned. At each level:
  \begin{enumerate}
  \item{Recompute the cluster matrices $U_A$ and $U_B$ for the regions $A$ and $B$ on either side of the separator.
   This computation uses as a reference configuration all spins $s_i=+1$, except on the boundary of the region $A\cup B$, where
   the fixed boundary spins are used as the reference configuration.}
   \item{Merge $U_A$ and $U_B$ using the edge weights for edges $e\in W(A,B)$ between $A$ and $B$  to obtain the
   matrix $K_W$, the effective Kasteleyn matrix for the separator $W(A,B)$.}
   \item{Compute $K_W^{-1}$ by Pfaffian elimination.}
   \item{For each of the $|W|-1$ edges $e\in W$, $e=(i,j)$ for a node $i$ on the boundary of $A$ and a node $j$ on the boundary of $B$:}
   \begin{enumerate}
     \item{Compute the probability of choosing $e$, $P(e)=\left|w_{ij}[K_W^{-1}]_{ij}\right|$.}
     \item{Choose whether to accept or reject the inclusion of $e$.}
     \item{Based on whether $e$ is included or excluded from the dimer sampling, set up the vectors $u$ and $v$ and apply the
        Sherman-Morrison formula to update $K_W^{-1}$.}
   \end{enumerate}
   \item{Fix the Ising spins that lie in the separator $W$ using the newly computed portions of the domain walls.}
   \end{enumerate}
\end{enumerate}

\section{Application and timing}

As the implementation of these methods into working programs is relatively complex,
we have carried out a number of tests of the code to confirm that it
computes partition functions and correlation functions correctly. Previous to this current version of the code,
each of the authors has independently written a computer code that computes
partition functions using Pfaffians. We confirmed
that the two previous codes and the current partition function code \cite{code} compute
the same partition function for samples of sizes up to size $L=256$, for several samples at each size.
We have also verified our code by comparison with (1) exact enumeration for 
small Ising spin glass samples of size up to $5\times 5$ spins and (2) checking correlation functions against analytic results for
the ferromagnetic Ising model.
The following subsections
summarize the results of tests for the pure Ising model and for spin glass models.
Similar checks are included as samples in our current distribution of the partition function code \cite{code}.
For further examples of applications of these particular codes, see
Refs.~\cite{ThomasHuseMiddleton,ThomasMiddletonSampling,ThomasHuseMiddletonUnpub,reentrance}.

\subsection{Verification in small samples}

The checks against exact enumeration verified that the code produced both correct partition functions and correlation functions.
A simple exact enumeration code computed the Boltzmann factor for each spin configuration $S$ directly, for a given
random selection of bond strengths $J_{ij}$. 
The sum of the Boltzmann factors at a given $\beta$ was compared against the partition function $Z$ computed
for each sample using nested dissection and multi-precision arithmetic. In all cases ($10^4$ random samples for each distribution), the
partition functions were in exact agreement.
We carried out tests for both bimodal and Gaussian distributions for $J_{ij}$.
As the support for the density of states is limited in the bimodal case when $J_{ij}=\pm 1$,
the bimodal distribution allows for an easy exact check of the number of states at each energy. 
Setting $\beta$ so that $\exp{\beta}=10^{m}$ for, say, $m=8$ allows one to directly read off the density of states from
$Z$ written in decimal, when the degeneracy at all energies is less than $10^{2m}$.

The spin-spin correlations generated by the nested dissection code were
also compared with the exact enumeration results and found to be the same.
In addition, to check our sampling code, up to $10^6$ configurations
were generated using our sampling methods for several $L=5$ samples.
The temperatures used were set so that about 90\% of the configurations were in one of the 
ten lowest energy states. The temperature was set this low to have enough statistics to verify the Boltzmann distribution for the low-lying states, including the degeneracies of the bimodal distribution.
The distribution of energies found were also found
to satisfy the Boltzmann distribution for Gaussian disorder, where there is a unique state for each energy.

\subsection{Verification of correlations using the ferromagnetic model}

\begin{figure}[h]
\centering
\includegraphics[width=2.8in]{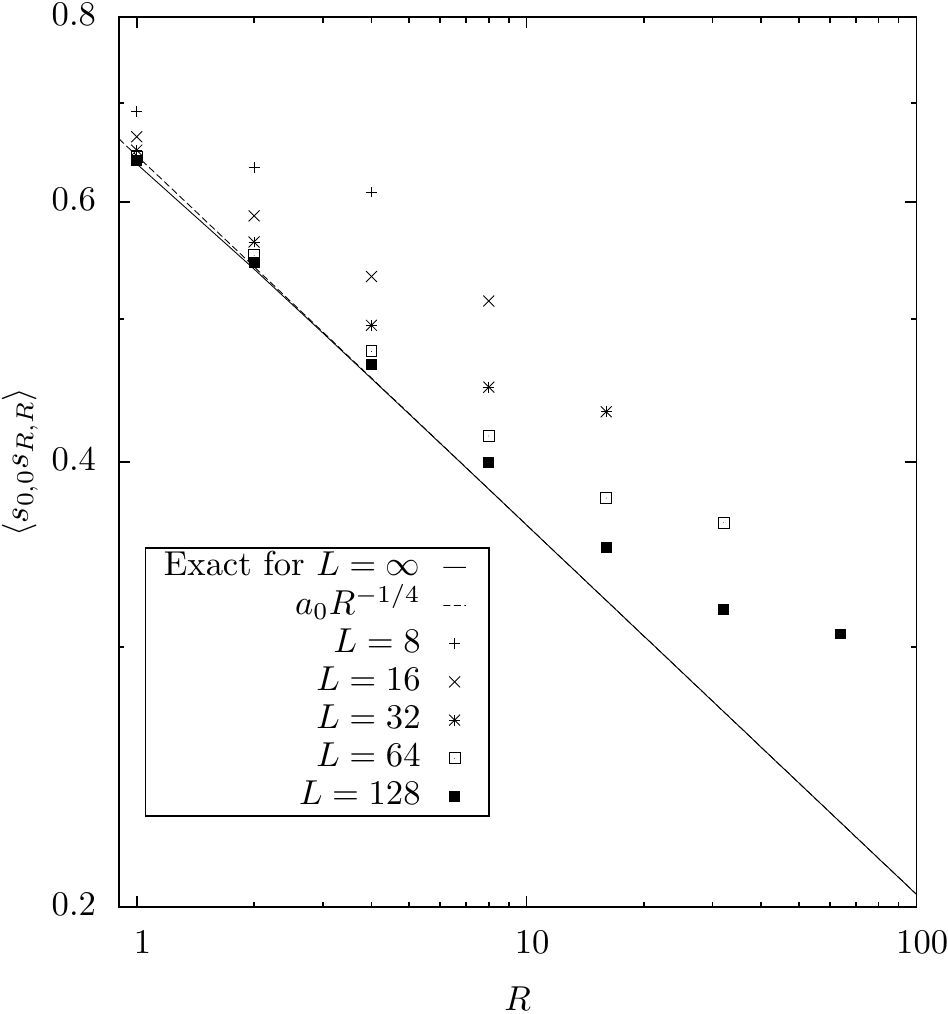}
\caption{Plot of correlation functions for the ferromagnetic Ising model at criticality. The correlation
function $\langle s(0,0)s(R,R) \rangle$ for diagonal spin separations $(R,R)$ was computed using Pfaffian methods for samples of size $L\times L$ spins, $L=8, \ldots, 128$, and compared with $L=\infty$ exact and asymptotic results.}
\label{fig:pure}
\end{figure}

To check the calculation directly against an analytic result in larger samples, we numerically computed
the spin-spin correlation function for the square lattice in ferromagnetic Ising models at
the critical temperature. For $J_{ij}=1$ for all neighboring pairs on the square the lattice, the critical
temperature $T_c$ satisfies $T_c^{-1}=\beta_c = \frac{1}{2}\ln(\sqrt{2}+1)$ for $J_{ij}=1$.
The correlation function along the diagonals, $\langle s_{0,0} s_{n,n}\rangle$, where the spins are now indicated
by two subscripts that indicate their $x$ and $y$ coordinates on the lattice.
The computed correlation function was compared with
known results \cite{60sCorrFn,ChengWu}. While this does not check for the effect of heterogeneities on
correlation functions, it helps confirm that correlation calculations are carried out correctly at all scales, from single
cities up through the size of the sample. The analytic result \cite{ChengWu}
for an infinite sample is
\begin{equation}\label{eqn:isingCorrFS}
\langle s_{0,0} s_{R,R} \rangle = \left(\frac{2}{\pi}\right)^R\Pi_{i=1}^{R-1}\left[1-\frac{1}{4i^2}\right]^{i-R}
\end{equation}Z
which at large separations $|i-j|=R\sqrt{2}\gg 1$ gives
\begin{equation}\label{eqn:isingCorrInf}
\langle s_{0,0} s_{R,R} \rangle = a_0 |i-j|^{-1/4}\,,
\end{equation}
with $a_0=2^{1/12}e^{3\zeta'(-1)}$ and $\zeta$ is the Riemann $\zeta$ function.
The numerical results for finite-size samples are plotted in \figref{fig:pure}. The rather large
finite-size corrections to the spin-spin
correlation functions are apparent, but the numerical calculation quickly converges to the exact short
distance results of \myeqref{eqn:isingCorrFS} and apparently converges to the asymptotic limit \myeqref{eqn:isingCorrInf},
giving us further confidence in the correlation function code.

\subsection{Timings}

We conclude the review of these algorithms with a list of the timings and memory used, in order
to compare with other implementations and algorithms. Table \ref{table:timing} gives average run
times on a single core of a 2.4 GHz Xeon E5620 quad-core processor with 12 GB of memory.
The GNU multiprecision arithmetic library
{\tt gmp} (version 3.5.0) was utilized for high precision floating point arithmetic
using the {\tt C++} interface {\tt gmpxx} (version 4.1.0) included with {\tt gmp} \cite{GMP}.
Multiprecision arithmetic was used for
all floating point calculations, including the temperature parameters, storing the bond strengths and weights,
and all matrix and partial Pfaffian operations. A pair of custom routines were written for the logarithm and
exponential functions. These are used only for computing weights at the start of the computation
and computing logarithms of the partition
functions at the end of the calculation, for finding free energies.
These timings are all for periodic samples: for samples with free or fixed
boundary conditions, approximately four times less memory and CPU time are needed.  As long as there
are no overflows, the running time is independent of inverse temperature $\beta$, though increasing the
precision increases the maximum value of $\beta$
for which the calculations are stable. For $L=256$ and bimodal disorder, floats using 1536 bits are needed to
reliably sample configurations 
for $\beta=20$. Computing the partition function or using Gaussian disorder requires just somewhat fewer bits.

\begingroup
\squeezetable
\begin{table*}[t!]
\begin{ruledtabular}
\begin{tabular}{|c|c|c|c|c|c|c|c|}
System size & Floating point & CPU time, & Peak memory, & CPU time, & Peak memory,  & CPU time, choose & Peak memory, choose\\
$L$  & precision (bits) & compute $Z$ & compute $Z$ & correlations  & correlations & configuration &
configuration \\
16 & 128 & 0.21 s & & 3.0 s& & 0.47 s& \\
16 & 512 & 0.45 s & & 3.9 s & & 0.83 s& \\
16 & 2048 & 2.9 s& & & & 4.75 s & 5.1 MB\\
32 & 128 & 1.0 s& & 18 s & 13 MB & 3.7 s& 4.9 MB\\
32 & 512 & 2.2 s & & 33 s& 17 MB & 6.3 s& 6.8 MB\\
32 & 2048 & 15 s & 11 MB & & & 37.1 s & 37 MB\\
64 & 128 & 5.9 s & 10 MB & 173 s & 47 MB & 29.0 s& 15 MB \\
64 & 512 & 13 s & 15 MB & 311 s & 67 MB & 50.4 s&  45 MB \\
64 & 2048 & 85 s & 38 MB & & &  301 s &  78 MB\\
128 & 128 & 40 s & 35 MB & 1873 s & 200 MB & --- &  ---\\
128 & 512 & 82 s & 57 MB & 2888 s & 281 MB &    348 s  & 87 MB\\
128 & 2048 & 552 s & 146 MB & & & 2454 s  & 222 MB \\
256 & 128 & 304s & 135 MB & & &  --- & ---\\
256 & 512 & 605 s & 224 MB & & & 3418 s & 240 MB\\
256 & 2048 & 3946 s & 580 MB & & & 20268 s & 899 MB
\end{tabular}
\end{ruledtabular}
\caption{A listing of timings and peak memory used by the Pfaffian nested dissection algorithms for different
calculations. The resource usage shown is for the computation for a single sample defined by
bond strengths $J_{ij}$. Timings and memory usage
are displayed for computing partition functions $Z$, for computing correlation
functions between all spins at the corners of the geometric dissection, and for choosing a single configuration
sampled exactly from the Boltzmann distribution.
The results are listed as a function of the size of the
square sample, with $N=L^2$ spins, and the floating point precision used. See the text for a brief description
of the hardware and multi-precision arithmetic libraries that were used. Blank entries in this table indicate where measurements were not made; dashes indicate where the matrix inverse used in the sampling method was unstable.}
\label{table:timing}
\end{table*}
\endgroup

\section{Future work}

The goal of this paper has been to present in detail numerical methods for computing thermodynamic
quantities, computing spin-spin correlation functions, and sampling configurations for two-dimensional Ising models with
short range (planar) interactions. The development and explication of these methods, which incorporates many ideas from previous
work, emphasizes the natural summing over various length scales. Especially at lower temperatures,
the near cancellations that result during the
matrix operations require matrix operations with multi-precision
arithmetic. The precise numerical results obtained
are a great advantage for studying thermodynamic quantities, compared with traditional Markov
Chain Monte Carlo methods, for a broad range of problems. We have prepared a code that should be easily
compiled to compute partition functions for the Ising model with arbitrary couplings on square lattices.
We are currently preparing implementations of the correlation function and sampling codes for
distribution. The structure of the algorithm is also very suggestive with respect
to the renormalization of couplings in random models, which might be studied directly to look for some type
of fixed point distribution in coarse grained couplings.

We thank Cris Cecka for introducing us to the FIND algorithm of Ref.\ \cite{FIND}. This work was
supported in part by the National Science Foundation grant DMR-1006731.  We thank the Aspen Center
for Physics, supported by NSF grant 1066293, for its hospitality while portions of this paper were written up.


\begin{thebibliography}{10}
\bibitem{footnote_a} This cancellation can most directly be understood
in terms of Fisher cities. The Fisher mapping from the Ising model to a dimer
model is one-to-one, and if one performs Pfaffian elimination on the two inner
nodes of a Fisher city, a Kasteleyn city results.  The two different types of
decorations are equivalent, forcing this convenient cancellation.
\end{thebibliography}
\end{document}